\documentclass[journal=nalefd,manuscript=letter,layout=twocolumn]{achemso}
\pdfoutput=1
\usepackage{graphicx}
\usepackage[utf8]{inputenc}
\usepackage{amsmath}
\usepackage{braket}
\usepackage{lmodern}
\usepackage[normalem]{ulem}
\usepackage{bbold}
\usepackage[T1]{fontenc}
\usepackage{xcolor}
\usepackage{siunitx}
\usepackage{hyperref}
\hypersetup{pdfnewwindow=true,allcolors=black,colorlinks=true}
\usepackage{lineno}

\newcommand {\nbs}{NbS$_2$}
\newcommand {\nbss}{Nb$_2$S$_3$}
\newcommand {\sphase}{$\sqrt{3}\times\sqrt{3}$ - phase}
\newcommand {\ophase}{$1\times1$ - phase}

\def\UK{%
	II. Physikalisches Institut,
	Universit\"at zu K\"oln,
	Z\"ulpicher Stra\ss e 77,
	D-50937 Cologne,
	Germany}

\def\LU{%
	NanoLund and Division of Synchrotron Radiation Research,
	Department of Physics,
	Lund University,
	SE-22100 Lund,
	Sweden}

\def\MAX{%
    MAX IV Laboratory,
    Lund University,
    SE-221 00 Lund,
    Sweden}

\def\FZJ{%
	Peter Gr\"{u}nberg Institut (PG-1),
    Forschungszentrum J\"{u}lich,
    Wilhelm-Johnen-Stra\ss e,
    D-52428 J\"{u}lich,
    Germany}

\title{Engineering two-dimensional materials from single-layer NbS$_2$}

\author{Timo~Knispel}
\affiliation\UK

\author{Daniela~Mohrenstecher}
\affiliation\UK

\author{Carsten~Speckmann}
\affiliation\UK

\author{Affan~Safeer}
\affiliation\UK

\author{Camiel~van~Efferen}
\affiliation\UK

\author{Virg\'{i}nia~Boix}
\affiliation\LU

\author{Alexander~Gr\"uneis}
\affiliation\UK

\author{Wouter~Jolie}
\affiliation\UK

\author{Alexei Preobrajenski}
\affiliation\MAX

\author{Jan~Knudsen}
\affiliation\LU
\affiliation\MAX

\author{Nicolae~Atodiresei}
\affiliation\FZJ

\author{Thomas~Michely}
\affiliation\UK

\author{Jeison~Fischer}
\email{jfischer@ph2.uni-koeln.de}
\affiliation\UK

\date{\today}

\begin{document}
\DeclareGraphicsExtensions{.pdf}

\begin{tocentry}
\includegraphics{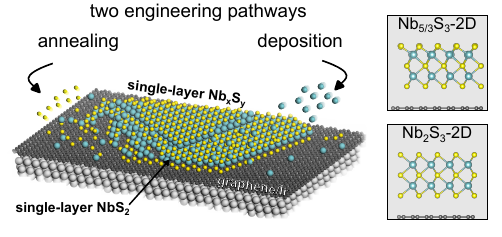}
\end{tocentry}

\noindent\begin{minipage}{\textwidth}
\begin{abstract}
Starting from a single layer of NbS$_2$ grown on graphene by molecular beam epitaxy, the single unit cell thick 2D materials Nb$_{5/3}$S$_3$-2D and Nb$_2$S$_3$-2D are created using two different pathways. Either annealing under sulfur-deficient conditions at progressively higher temperatures or deposition of increasing amounts of Nb at elevated temperature result in phase-pure Nb$_{5/3}$S$_3$-2D followed by Nb$_2$S$_3$-2D. The materials are characterized by scanning tunneling microscopy, scanning tunneling spectroscopy and X-ray photoemission spectroscopy. The experimental assessment combined with systematic density functional theory calculations reveals their structure. The 2D materials are covalently bound without any van der Waals gap. Their stacking sequence and structure are at variance with expectations based on corresponding bulk materials highlighting the importance of surface and interface effects in structure formation. 

Keywords: niobium disulfide, single layer, molecular beam-epitaxy, covalent transformation
\end{abstract}
\end{minipage}

\clearpage

\newpage
\noindent
Thinning down a layered material to few or single layers transforms it into a two-dimensional (2D) material. The typical way of obtaining a 2D material is by exfoliation of the bulk crystal. The method is simple, the structural quality of exfoliated layers is generally very good \cite{Huang15,Huang20}, stacking of layers to create vertical heterostrutures with new functions is straightforward \cite{Geim13,Novoselov16}, and finally twisted stacking opened the door for moir\'{e} physics \cite{Cao18}. 

Nevertheless, exfoliation as a method has several significant limitations. Beyond its fundamental scalability issues, exfoliation is ineffective in preparing single or few-layer thick 2D materials from covalently bound bulk crystals. Such crystals lack van der Waals gaps and, consequently, cannot be adequately exfoliated. Additionally, exfoliation cannot be used for synthetically constructed 2D materials that have no bulk counterparts in terms of structure or composition.

The scope of 2D materials can be substantially broadened by the use of growth methods like molecular beam epitaxy (MBE) or chemical vapor deposition (CVD). For example, provision of more than one metal during growth enables one to explore the entire composition space between two dissimilar transition metal dichalcogenides (TMDCs) on the single layer level \cite{Wan22} or to create vertical TMDC heterostructures with continuously tunable moir\'{e} periodicity by using the composition dependent lattice parameters \cite{Fortin24}. Variation of the metal chemical potential enables the production of thin films across the entire sequence of self-intercalation compounds known from bulk crystals, allowing for the discovery of magnetic order in some of these intercalated phases \cite{Zhao20}.

Annealing an initial transition metal chalcogenide with or without chalcogene flux, possibly following prior metal deposition, is another strategy applied to create new phases. Examples are the annealing-induced single-layer transformations of CrSe$_2$ into Cr$_2$Se$_3$ \cite{Liu21}, of VS$_2$ into stripped V$_2$S$_3$ \cite{Arnold18} or V$_4$S$_7$ \cite{Efferen24}, of PtTe$_2$ into Pt$_2$Te$_2$ \cite{Lasek22}, of $\alpha$-FeSe into kagome Fe$_5$S$_8$ \cite{Zhang23}, or of Bi$_2$Se$_3$ into MnBi$_3$Se$_4$ \cite{Khatun24}. The enumeration is by far not complete.

In the present manuscript we investigate phase transitions of single-layer NbS$_2$. This TMDC has attracted substantial research interest due to its superconductivity in the bulk \cite{vanmaaren66,Fisher80,Witteven21} and its charge-density wave in the single layer \cite{Lin18,Knispel24}. Moreover, self-intercalated Nb$_\mathrm{1+x}$S$_2$ is an excellent catalyst for the hydrogen evolution reaction~\cite{Yang19}.

Besides the van der Waals material NbS$_2$, the Nb-S phase diagram displays a zoo of phases without van der Waals gap, \textit{i.e.}, being covalently bound, of which the structures were carefully investigated by X-ray diffraction \cite{Jellinek60,Kadijk69}. Whether any of these covalently bound bulk phases possesses a 2D pendant is still unexplored.    
Here we establish and characterize two Nb$_x$S$_y$-2D compounds of single unit cell thickness, namely Nb$_{5/3}$S$_3$-2D and Nb$_2$S$_3$-2D. To avoid confusion with bulk materials with the same composition, but different structure, "-2D" is attached to the stoichiometric formulas indicating the yet undescribed 2D materials. Due to their single unit cell thickness, these compounds are referred to as single-layer materials. A single layer consists of stacked S-Nb-S-Nb-S planes of atoms. Starting from MBE-grown single-layer NbS$_2$, the new phases are established under ultrahigh vacuum conditions via two different kinetic pathways, either through pure annealing or by metal deposition at elevated temperature. We find that each phase can be prepared phase-pure, making its investigation by averaging techniques feasible.

\begin{figure*}[h!]
	\centering
	\includegraphics[width=0.7\textwidth]{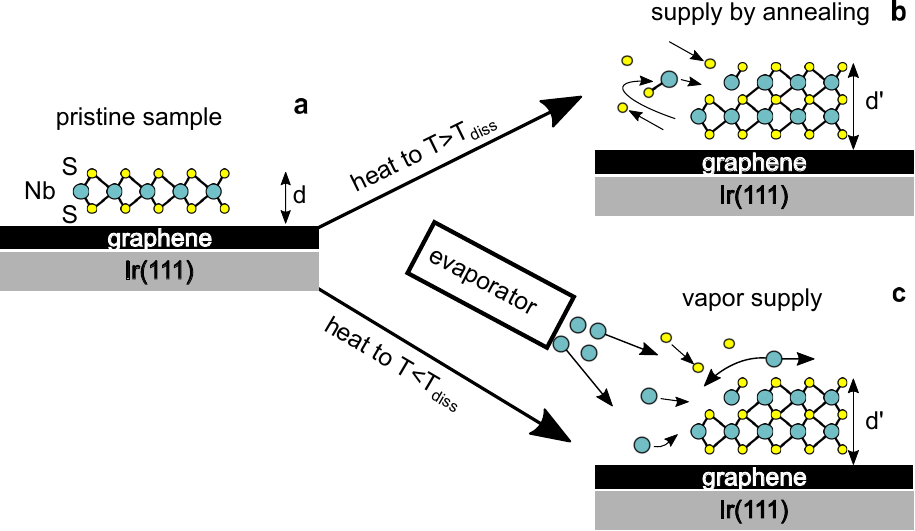}
	\caption{Concept of covalent transformation. (a) Single-layer H-\nbs~on Gr/Ir(111). (b) Covalent transformation by heating and dissociation. (c) Covalent transformation by deposition of additional Nb.}
	\label{covalent_transformation_concept}
\end{figure*}

Beyond our methodology for creating these materials, we highlight three important findings that are of broader relevance and can be generalized to the covalent growth of other layered 2D materials. Firstly, applying careful titration based on a well calibrated evaporator is an efficient tool to determine the stoichiometry of an unknown compound resulting from phase transformation. Secondly, the structures of the resulting 2D materials differ from the known bulk phases, although their chemical composition is rather similar. Our results, thus indicate that surface effects are important and consequently assumptions that the compounds can be described from corresponding bulk phases may fail. Thirdly, we show that the explicit inclusion of the substrate in theoretical calculations is necessary to provide a valuable insight into the range of possible phases and to guide the interpretation of experiments.

\section*{Results}
\subsection*{Concepts for covalent transformation of single-layer \nbs}
\noindent
Our course of action to achieve covalent growth is exemplified for \nbs~in Figure~\ref{covalent_transformation_concept}. Figure~\ref{covalent_transformation_concept}a displays single-layer H-\nbs~grown on graphene (Gr) on Ir(111). The first strategy is to heat up the sample to a temperature $T_\mathrm{diss}$ that causes \nbs~to partially dissociate [Figure~\ref{covalent_transformation_concept}b]. Some of the S of \nbs~escapes into vacuum or intercalates between Gr and Ir(111). The remaining Nb excess triggers a phase transformation to a covalently bonded niobium-rich compound composed of 3 atomic planes of S separated by Nb planes. The new compound exhibits an increased height $d'$. Evidently, the partial dissociation of the single-layer \nbs~of height $d$ in combination with the larger height $d'$ of the new phase formed without supply of additional material causes a reduction in sample coverage. The second strategy is to induce covalent growth by deposition of additional Nb at a temperature $T<T_{diss}$, see Figure~\ref{covalent_transformation_concept}c. When arriving on the surface, the deposited Nb reacts with the existing \nbs~and thereby triggers the phase transformation. In both cases identical phases can be obtained.

\subsection*{Covalent transformation by N\lowercase{b}S$_2$~annealing}
\noindent
This section describes the transformation of \nbs~into two different phases richer in Nb obtained by heating to successively higher temperatures with $T > T_{diss}$.

\begin{figure*}[h!]
	\centering
	\includegraphics[width=0.9\textwidth]{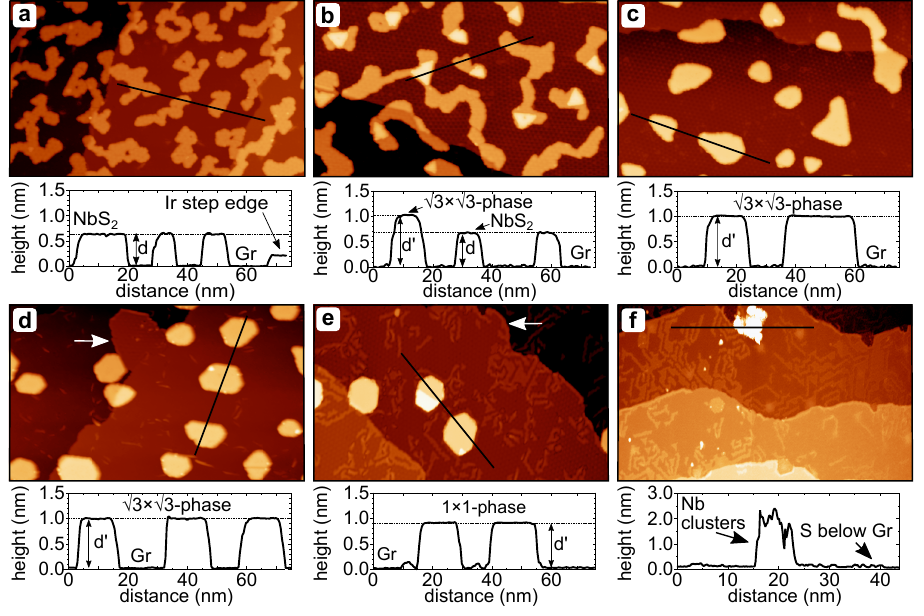}
	\caption{STM topographs of an isochronal annealing sequence of initial single-layer \nbs~islands without supply of additional S. Annealing intervals are 360\,s. (a) single-layer \nbs~islands after room temperature growth and annealing to 820\,K. (b)-(f) After additional annealing to (b) 920\,K, (c) 1020\,K, (d) 1120\,K, (e) 1220\,K, and (f) 1320\,K. Height profiles along the black lines are shown below the topographs. Height levels $d=0.62$\,nm, $d'=0.99$\,nm and $d'=0.93$\,nm distinguish between single-layer \nbs, \sphase, \ophase, respectively. Image information: for all size $\mathrm{150\,nm~ \times}~90$\,nm, a $V_\mathrm{s}=$1.0\,V,  $I_\mathrm{t}=0.23$\,nA; (b) $V_\mathrm{s}=0.95$\,V, $I_\mathrm{t}=0.34$\,nA; (c) $V_\mathrm{s}=1.0$\,V, $I_\mathrm{t}=0.26$\,nA; (d) $V_\mathrm{s}=0.92$\,V, $I_\mathrm{t}=0.33$\,nA; (e) $V_\mathrm{s}=1.0$\,V, $I_\mathrm{t}=0.32$\,nA; (f) $V_\mathrm{s}=2.2$\,V, $I_\mathrm{t}=0.06$\,nA.}
	\label{STM_anneal_seq}
\end{figure*}

Figure~\ref{STM_anneal_seq}a displays a scanning tunneling microscopy (STM) image of pristine single-layer \nbs~islands grown on Gr/Ir(111) by room temperature deposition of 0.34\,ML Nb in S vapor and subsequent annealing to 820\,K (compare Methods). The islands cover an area fraction of 0.34, are continuous over Ir substrate steps under the Gr carpet, and are of irregular shape. The islands display an apparent height of $d=0.62\pm0.01$\,nm at $V_\mathrm{s}=1.00$\,V as exemplified by the height profile below the topograph ($d$ depends slightly on the tunneling voltage; $d=0.58\pm0.01$\,nm at $V_\mathrm{s}= - 1.00$\,V). The measured apparent heights fit reasonably well to the apparent height of 0.578\,nm reported for single-layer \nbs~on Gr/6H-SiC(0001)~\cite{Lin18} and to half of the c-axis lattice constant of 1.195\,nm of bulk \nbs~\cite{Fisher80}. In our previous work \cite{Knispel24}, it was firmly established that under these conditions \nbs~on Gr/Ir(111) grows in the H-phase, in agreement with the findings for single-layer \nbs~on Au(111) \cite{Stan19} and bulk \nbs~\cite{Fisher80}.

While \nbs~islands are stable at 820\,K independent of the duration of annealing, after annealing the sample to 920\,K, the island area fraction decreases to 0.29. Higher triangular shaped areas emerge within the \nbs~islands [Figure~\ref{STM_anneal_seq}b]. In these areas, height profiles give an increased apparent height of $d'=0.99$\,nm at $V_\mathrm{s}=1.00$\,V ($d'=0.90$\,nm at $V_\mathrm{s}= -1.00$\,V). This height is inconsistent with bilayer \nbs, which has an apparent height of $1.22$\,nm at $V_\mathrm{s}=1.00$\,V, as displayed in Figure~S1 in the SI. These higher areas are designated as in the \sphase, since below it will be shown that they exhibit a $(\sqrt{3}\times\sqrt{3})\mathrm{R}30^\circ$ superstructure and are of composition Nb$_{5/3}$S$_3$-2D.

After annealing the sample to 1020\,K [Figure~\ref{STM_anneal_seq}c], all islands display an apparent height of $d'=0.99$\,nm, consistent with the assumption that the islands have entirely transformed to the \sphase, indicating phase purity. The island area fraction decreased further to 0.17.  

At first glance, annealing the sample to 1120\,K as shown in Figure~\ref{STM_anneal_seq}d does not change the situation. From the height profile, all islands still display a height $d'=0.99$\,nm characteristic of the \sphase. The islands shape is more regular, mostly hexagonal. However, the island area fraction further decreases to 0.12 and new features appear as highlighted by the white arrow. It marks a peninsula attached to an Ir step edge. This and other peninsulas are attributed to Nb that penetrated the Gr sheet and attached to a step on Ir(111). Indeed, metal deposited or liberated from a TMDC layer on Gr/Ir(111) intercalates at elevated temperatures \cite{Schumacher13,Galera17,Hall18}.

The \sphase~islands are easy to shift laterally as a whole by the STM tip (see Figure~S2a in the SI) consistent with being physisorbed to Gr. The liberated Gr, initially under an island, displays neither structural nor height changes. Thereby, it is ruled out that the height increase to $d'=0.99$\,nm is caused by changes in the Gr height level, \textit{e.g.}, by intercalation of S or Nb underneath the islands.

After annealing to 1220\,K [Figure~\ref{STM_anneal_seq}e] the island area fraction is further reduced to 0.10. Two different height levels are present, neither of which coincides with the aforementioned ones. As obvious from the height profile, the dominant height level is $d'=0.93$\,nm at $V_\mathrm{s}=1.00$\,V ($d'=0.94$\,nm at $V_\mathrm{s}= -1.00$), slightly, but clearly, lower than the \sphase~height. We designate areas with this height level as \ophase, since below it will be shown that they exhibit no superstructure and are of composition Nb$_{2}$S$_3$-2D.

In addition to Nb intercalation peninsulas attached to Ir steps (white arrow) already observed after annealing to 1120\,K, stripes and patches with an apparent height of about 0.1\,nm with respect to the Gr base level are prominent now. In LEED this new intercalation feature gives rise to  $(\sqrt{3}\times\sqrt{3})\mathrm{R}30^\circ$ spots with respect to Ir(111) [compare Figure~S3 in the SI]. Sulfur is well known to form a $(\sqrt{3}\times\sqrt{3})\mathrm{R}30^\circ$ superstructure on Ir(111) and between Ir(111) and Gr~\cite{Pielic20}. Thus, these stripes and patches are due to S intercalation between Gr and Ir(111). A few of these intercalation stripes are already present at lower annealing temperature [compare Figure~\ref{STM_anneal_seq}d]. Sulfur intercalation between Gr and Ir(111) is similarly observed after growth and annealing of other TMDCs~\cite{Hall18}.

The final annealing step to 1320\,K leads to the complete decomposition of \nbs~and no more Nb-S islands are visible, see Figure~\ref{STM_anneal_seq}f. What remains is metallic Nb clusters as well as intercalated S and Nb.    

\begin{figure*}[h!]
	\centering
	\includegraphics[width=0.9\textwidth]{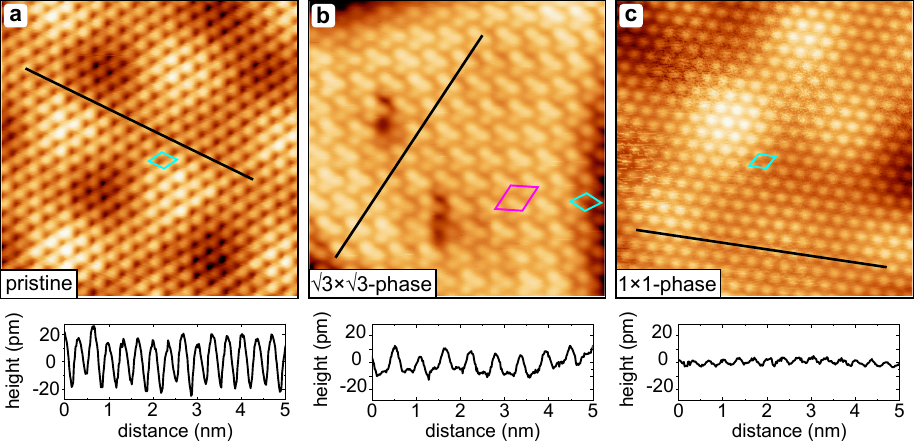}
	\caption{Atomic resolution STM topographs of (a) pristine single-layer \nbs, (b) the \sphase, and (c) the \ophase~taken at 1.7\,K. In the STM topographs the unit cells of the three phases are indicated by cyan rhomboids. Magenta rhomboid is the $(\sqrt{3}\times\sqrt{3})\mathrm{R}30^\circ$ superstructure. Height profiles along the black lines are shown below the topographs. Image information: for all size $\mathrm{6\,nm\times6}$\,nm and $T_\mathrm{s} = 1.7$\,K, (a) $V_\mathrm{s}=50$\,mV, $I_\mathrm{t}=0.5$\,nA; (b) $V_\mathrm{s}=100$\,mV, $I_\mathrm{t}=0.80$\,nA; (c) $V_\mathrm{s}=100$\,mV, $I_\mathrm{t}=0.70$\,nA.} 
	\label{STM_phases}
\end{figure*}

To justify the designation of the 0.99\,nm or 0.93\,nm high islands as being \sphase~or \ophase, we present atomically resolved STM images of the island structures from Figures~\ref{STM_anneal_seq}a, c and e. For reference, Figure~\ref{STM_phases}a displays atomically resolved \nbs. It has a lattice parameter of 0.331(3)\,nm as established in our previous work \cite{Knispel24}, in good agreement with values found for the single-layer \nbs~on bilayer Gr/6H-SiC(0001) (0.334\,nm)~\cite{Lin18} and bulk 2H-\nbs~(0.3324\,nm)~\cite{Fisher80}. The atomic corrugation is on the order of 35\,pm as apparent from the height profile along the black line shown below the topograph. The height modulation on a length scale of $\approx 2.5$\,nm is due to the Gr/Ir(111) moir\'{e} pattern imposed on \nbs~\cite{Hall18,Knispel24}. 

An atomically resolved topograph of the \sphase~is shown in Figure~\ref{STM_phases}b. It displays a clear $(\sqrt{3}\times\sqrt{3})\mathrm{R}30^\circ$ superstructure with respect to the original \nbs~lattice. tm{Since all atoms in the top layer can still be recognized,
the superstructure results from a trimerization of the S atoms.} At the island edges the trimerization seems to fade away. The superstructure is associated with a height modulation on the order of 15\,pm. The superstructure displays a lattice size of $0.577\pm0.05$\,nm, which corresponds to $\sqrt{3}a$, with $a=0.333\pm0.03$\,nm. LEED obtained after annealing to $1020$~K displays faint spots of a $(\sqrt{3}\times\sqrt{3})\mathrm{R}30^\circ$ superstructure with respect to \nbs~[Figure~S3 in the SI].

Figure~\ref{STM_phases}c shows an atomically resolved topograph of the \ophase. It features a $ 1 \times 1$ structure with the lattice parameter $a=0.330\pm0.05$\,nm, identical to the one of single-layer \nbs~within the limits of error. The \ophase~is distinct from pristine \nbs~because: (i) it differs in height (0.93\,nm vs. 0.62\,nm); (ii) the atomic corrugation is only in the order of 5\,pm, reduced by about a factor of 7 compared to the pristine \nbs; and (iii) the Gr/Ir(111) moir\'{e} corrugation is shining through the \ophase~islands is considerably damped compared to \nbs.

\begin{figure*}[t]
	\centering
	\includegraphics[width=0.9\textwidth]{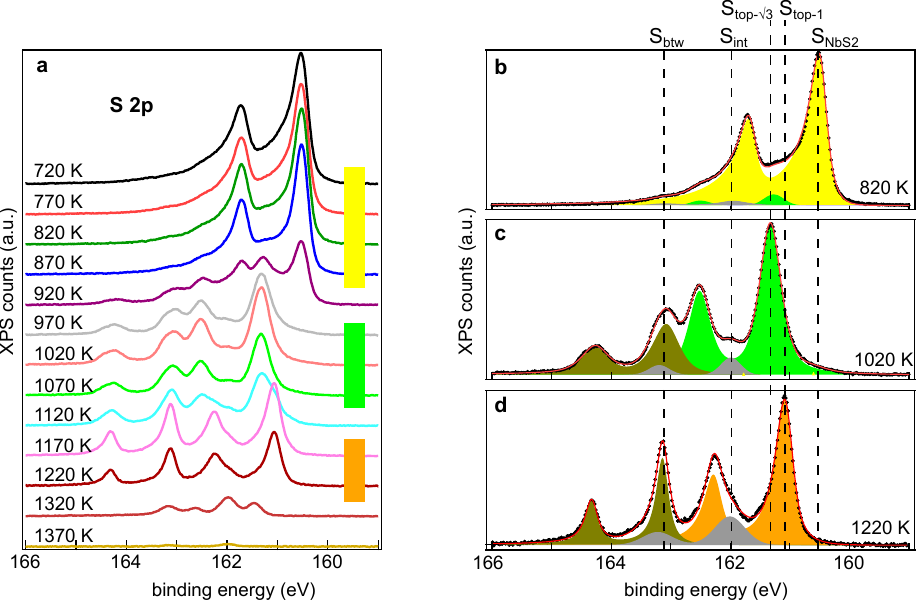}
	\caption{(a) XPS of the S~2p core level of initial single-layer \nbs~on Gr/Ir(111) transformed during annealing. After room temperature growth, for each spectrum the sample was annealed to the indicated temperature without supply of additional S and cooled down to 300\,K for measurements. The spectra are grouped in three temperature ranges according to their similarities: yellow, green, and orange. (b)-(d) S~2p core level spectra after annealing to (b) 820~K, (c) 1020~K, and (d) 1220~K fitted with components.
	}
	\label{xps_dissociation_S2p}
\end{figure*}

In order to obtain complementary chemical information about the annealing-induced phases, X-ray photoemission spectroscopy (XPS) of the S~2p, Nb~3d, C~1s, and Ir~4f core levels was performed. Figure~\ref{xps_dissociation_S2p}a shows the S~2p core-level spectra of samples with increasing annealing temperatures from 720~K (top) to 1370~K (bottom). The S~2p core level is spin-orbit split into a 2p$_{3/2}$ and 2p$_{1/2}$ doublet with an energy separation of $1.19(3)$~eV. The S~2p components are referenced in the following to the lower binding energy 2p$_{3/2}$ peak. The appearance of the spectra has three distinct temperature ranges: (i) 720~K up to 870~K (yellow bar), (ii) 970~K to 1070~K (green bar), and (iii) 1170~K to 1220~K (orange bar). These temperature ranges agree well with the temperature ranges of the phases identified in STM (compare Figure~\ref{STM_anneal_seq}). The distinct spectra for each temperature range is evidence of phase purity. The sequence of spectra shows a decrease in S~2p intensity with temperature (see also Figure~S4 in the SI), and after annealing at 1320~K the S~2p signal has nearly vanished consistent with the decomposition of Nb-S compounds and subsequent S desorption. Spectra for Nb~3d, C~1s, and Ir~4f core-levels are displayed in Figure~S5 of the SI.

The spectra in Figure~\ref{xps_dissociation_S2p}b,c,d correspond to single-layer \nbs, \sphase, and \ophase, respectively. The S~2p spectrum of \nbs~in Figure~\ref{xps_dissociation_S2p}b displays mainly a single spin-orbit doublet (yellow), designated S$_{\mathrm{NbS2}}$ and located at 160.60(0)~eV. The S$_{\mathrm{NbS2}}$ component is attributed to top and bottom S in \nbs. In the \sphase~spectrum after annealing to 1020\,K in Figure~\ref{xps_dissociation_S2p}c the S$_{\mathrm{NbS2}}$ component is absent and two new main components are present: S$_{\mathrm{top-\sqrt{3}}}$ at 161.33~eV (green) and S$_{\mathrm{btw}}$ at 163.08~eV (olive). 

The substantial core level shifts are consistent with a phase transformation from \nbs~to the \sphase. Having the largest intensity, the S$_{\mathrm{top-\sqrt{3}}}$ component is associated with the top sulfur layer. The S$_{\mathrm{btw}}$ component (olive) with a significant core level shift of $2.48$~eV compared to S$_{\mathrm{NbS2}}$ is tentatively assigned to sulfur in a lower atomic plane (due to its lower intensity) and in a very different chemical environment than in \nbs. The origin of the S$_{\mathrm{btw}}$ will be clarified further below with the help of additional information from STM and DFT. The gray $S_{\mathrm{int}}$ component at 161.99~eV is attributed to intercalated S lost during the phase transformation. The same S~2p component growing in intensity during annealing has been found for VS$_2$ \cite{Efferen24}.

The \ophase~spectrum after annealing to 1220\,K in Figure~\ref{xps_dissociation_S2p}d retains the S$_{\mathrm{btw}}$ component at 163.13\,eV (olive) and develops a new S$_{\mathrm{top-1}}$ component at 161.05~eV (orange), shifted by 0.28\,eV with respect to the S$_{\mathrm{top-\sqrt{3}}}$ component. Overall, the spectrum is quite similar to the \sphase~spectrum and distinct from the \nbs-spectrum. Remarkably, the S$_{\mathrm{top-1}}$ and S$_{\mathrm{btw}}$ components are sharp with full width at half maximum (FWHM) of 0.29~eV and 0.27~eV, almost halved compared to the \sphase. The narrow peaks indicate the homogeneous state of the S in this phase. The S$_{\mathrm{int}}$ component is further increased due to the release of sulfur during the transformation from the \sphase~to the \ophase.

\subsection*{Covalent transformation of N\lowercase{b}S$_2$ by Nb vapor supply}

\noindent
The phase transformations of \nbs~are likely to be triggered by Nb excess resulting from the loss of S due to annealing. If this rationale is correct, one could expect the transformation also to take place already at temperatures below $T_\mathrm{diss}$, if additional Nb is supplied. Moreover, by controlling the amount of Nb supplied, it might be possible to select the resulting phase.

\begin{figure*}[htb]
	\centering
	\includegraphics[width=0.6\textwidth]{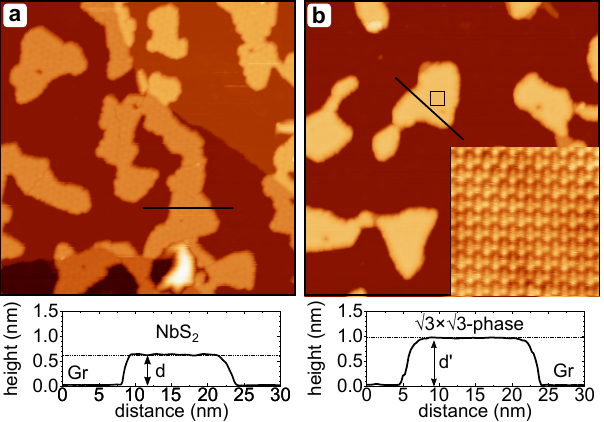}
	\caption{Formation of the \sphase~by Nb supply. (a) Single-layer \nbs~grown by deposition of 0.36\,ML Nb in S background pressure at room temperature and annealed to 820\,K in the absence of additional S supply. (b) Sample after deposition of additional 0.12 ML Nb at 820\,K in the absence of additional S supply. Inset: atomic resolution topograph of boxed area. Height profiles are taken along the black lines in the STM topographs. Image information: (a) size $\mathrm{100\,nm\times100}$\,nm, $V_\mathrm{s}=1.0$\,V, $I_\mathrm{t}=1.00$\,nA; (b) size $\mathrm{100\,nm\times100}$\,nm, $V_\mathrm{s}=1.0$\,V, $I_\mathrm{t}=1.0$\,nA; Inset: $\mathrm{5\,nm\times5}$\,nm, $V_\mathrm{s}=0.1$\,V, $I_\mathrm{t}=5$\,nA.}
	\label{STM_1_3_Nb_vapor}
\end{figure*}

To test this idea, 0.12\,ML Nb was deposited at 820\,K on pre-grown single-layer \nbs~islands with an area fraction of 0.36 [Figure~\ref{STM_1_3_Nb_vapor}a]. Plain annealing at 820\,K neither causes \nbs~dissociation nor changes the coverage fraction. Upon deposition of Nb, the single-layer \nbs~transforms into the \sphase~[Figure~\ref{STM_1_3_Nb_vapor}b]: the island height increased to 0.99\,nm and the atomic resolution inset displays a $(\sqrt{3}\times\sqrt{3})\mathrm{R}30^\circ$ superstructure. The island area fraction decreased from 0.36 to 0.28. This titration experiment allows one to calculate the amount of Nb per unit cell in the \sphase. The total amount of Nb provided consists of 0.36\,ML + 0.12\,ML while the island area fraction is 0.28. Thus, each \sphase~unit cell contains $x=\frac{0.36+0.12}{0.28}$ or $x=1.71$ Nb atoms within the limits of error. It appears likely that $x = 5/3$ due to a full Nb layer and an additional 2/3 Nb layer. Provided the 1/3 vacancies order, they could give rise to the $\sqrt{3} \times \sqrt{3}$ superstructure of the \sphase.

\begin{figure*}[htb]
	\centering
	\includegraphics[width=0.9\textwidth]{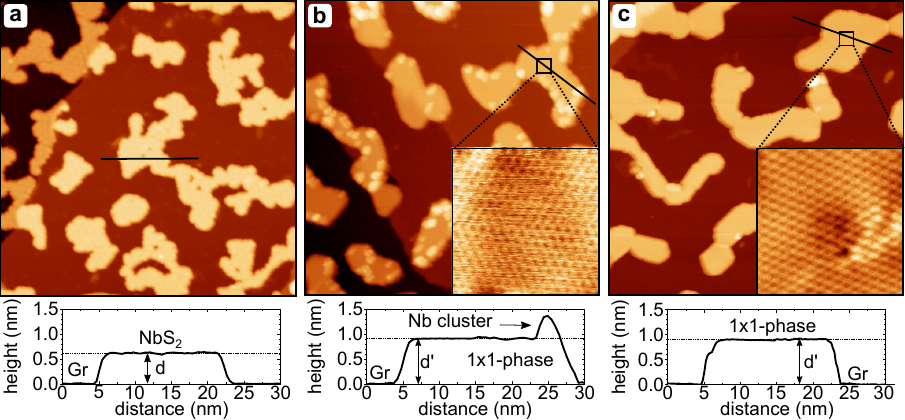}
	\caption{Formation of the \ophase~by Nb supply. (a) Single-layer \nbs~grown by deposition of 0.33\,ML Nb in S background pressure at room temperature and annealed to 820\,K in the absence of additional S supply. (b) Sample after deposition of additional 0.33\,ML Nb at 820\,K. Inset: atomic resolution topograph of the boxed area. (c) Sample after additional annealing to 1020\,K. Inset: atomic resolution STM topograph of the boxed area. Height profiles are taken along the black lines in the topographs. Image information: (a) size $\mathrm{100\,nm\times100}$\,nm, $V_\mathrm{s}=1.0$\,V, $I_\mathrm{t}=0.23$\,nA; (b) size $\mathrm{100\,nm\times100}$\,nm, $V_\mathrm{s}=1.0$\,V, $I_\mathrm{t}=0.3$\,nA; Inset: $\mathrm{5\,nm\times5}$\,nm, $V_\mathrm{s}=0.1$\,V, $I_\mathrm{t}=5$\,nA; (c) size $\mathrm{100\,nm\times100}$\,nm, $V_\mathrm{s}=1.2$\,V, $I_\mathrm{t}=0.3$\,nA; Inset: $\mathrm{5\,nm\times5}$\,nm, $V_\mathrm{s}=0.1$\,V, $I_\mathrm{t}=5$\,nA.}
	\label{STM_Nb_vapor}
\end{figure*}

Similarly, we deposited 0.33\,ML Nb at 820\,K on pre-grown pristine single-layer \nbs~islands with an area fraction of 0.33 [Figure \ref{STM_Nb_vapor}a]. As apparent from Figure~\ref{STM_Nb_vapor}b, upon deposition the \nbs~islands transformed to the \ophase: the island height increased to 0.93\,nm and the atomic resolution inset displays a $1\times1$ structure with low corrugation. Additionally, small clusters of large height are present at the island edges. The island area fraction  marginally decreased from 0.33 to 0.29. With the same approach as above, one obtains formally an Nb content of 2.3 atoms per \ophase~unit cell. We tentatively conclude that a \ophase~unit cell contains 2 Nb atoms while the excess Nb is contained in the metallic clusters. The sample remains in the \ophase~upon additional annealing to 1020\,K while the clusters at the island edges largely disappear [compare Figure~\ref{STM_Nb_vapor}c]. Their disappearance is presumably due to Nb already escaping under Gr at 1020\,K.

It is remarkable that the \ophase~forms at 820\,K when excess Nb is supplied, whereas plain annealing requires 1220\,K. The kinetically difficult process in phase formation during annealing under sulfur-poor conditions appears to be the release of Nb rather than reorganization of the initial \nbs~islands.

\begin{figure}[htb]
	\centering
	\includegraphics[width=0.45\textwidth]{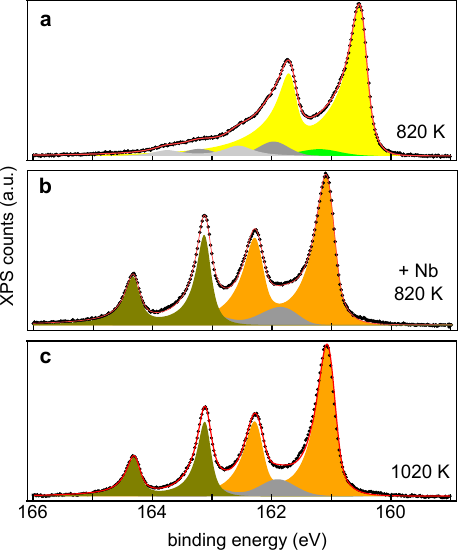}
	\caption{(a) High-resolution XPS of the S~2p core level of \nbs~on Gr/Ir(111). (b) After deposition of additional Nb at 820~K. (c) After annealing at 1020~K. Fit of each spectrum with five S~2p components.
	}
	\label{XPS_vapor_S2p}
\end{figure}

XPS corroborates the transformation of \nbs~to the \ophase~using the same conditions as for the STM sequence presented in Figure~\ref{STM_Nb_vapor}. Figure~\ref{XPS_vapor_S2p}a shows the typical S~2p spectrum for pristine \nbs, similar to Figure~\ref{xps_dissociation_S2p}b. After deposition of Nb at 820~K, the spectrum in Figure~\ref{XPS_vapor_S2p}b is nearly identical to the spectrum obtained after annealing to 1220~K in the absence of Nb supply [compare Figure~\ref{xps_dissociation_S2p}d] being characteristic of the \ophase. Upon further annealing to 1020~K, the \ophase~remains unchanged and the related spectrum in Figure~\ref{XPS_vapor_S2p}c is indistinguishable from the one obtained after \ophase~ formation at 1220\,K without additional Nb supply shown in Figure~\ref{xps_dissociation_S2p}d.

\subsection*{DFT calculations}
\noindent
With the experimental information at hand and the help of density functional theory (DFT) calculations we determined the structure of the 2D materials resulting from phase transformations of \nbs. We start the analysis with the \ophase. 

The \ophase~has the following properties: (i) hexagonal symmetry and lattice parameter identical to single-layer \nbs~within the limits of error; (ii) it contains a smaller fraction of sulfur than \nbs~and even less than the \sphase, since it evolves upon annealing from these phases under sulfur-deficient conditions, accompanied by a gradual decrease of the  S~2p intensity; (iii) apparent height 0.93\,nm, larger by 0.31\,nm compared to \nbs; (iv) 2 Nb atoms per unit cell; (v) no superstructure; (vi) only physisorbed to Gr.

\begin{figure}[h]
	\centering
	\includegraphics[width=0.45\textwidth]{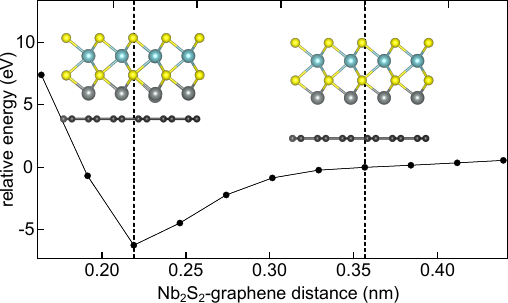}
	\caption{Chemisorption of Nb$_2$S$_2$-2D to Gr. Relative total energy of minimum energy configuration of Nb$_2$S$_2$-2D (lowest energy structure) as a function of the distance to Gr. Zero point of the energy scale is at 0.36\,nm in the physisorbed state. Inset: side view ball models of relaxed DFT geometries for Nb$_2$S$_2$-2D in the 0.36~nm~and 0.22~nm Nb-C distances.
 	Nb atoms: cadet blue balls and gray; S atoms: yellow; dark gray: C atoms.
	}
	\label{4-layers_z-dep}
\end{figure}

It is well known from the Nb-S phase diagram and previous reports \cite{Jellinek60,Kadijk69} that at high temperatures in bulk a NiAs-type structure of NbS forms, with almost identical lattice parameter as \nbs. The NiAs structure is hexagonal, with As atoms forming a hexagonal close-packed lattice, with Ni in octahedral sites, resulting in alternating atomic planes of Ni and As. Since the \ophase~contains two Nb atoms per unit cell, a natural starting point for the DFT calculations was just the NbS bulk unit cell, \textit{i.e.}, Nb$_2$S$_2$-2D.

The minimum energy configuration of Nb$_2$S$_2$-2D on Gr is surprisingly not of NiAs-type. It consists of two Nb layers in trigonal prismatic coordination with the S atoms as well as the Nb atoms of the respective layers sitting atop each other, as shown in the ball model insets of Figure~\ref{4-layers_z-dep}. For higher energy structures compare Table~S1 in the SI. The ground state configuration (i.e., the lowest energy structure) is chemisorbed to Gr with Nb$-$plane to Gr distance of only 0.22\,nm. However, our calculations also identified a local minimum configuration in which the Nb$_2$S$_2$-2D layer is physisorbed at a distance of 0.36\,nm. Furthermore, starting from the chemisorbed configuration and rigidly lifting the Nb$_2$S$_2$-2D above the Gr we evaluated the total energy of the system at specific distances along the $z-$direction. Figure~\ref{4-layers_z-dep} shows the total slab energy versus distance. No significant barrier exists between the physisorbed and the chemisorbed state. As an additional note, the chemisorption of hypothetical Nb$_2$S$_2$-2D to Gr is not surprising, given the expected high reactivity of the bare Nb (\textit{i.e.}, a $4d$ metal) towards the C atoms of Gr. Considering the high temperatures used in our experiments, one expects that any potential barrier between physisorbed and chemisorbed state can be overcome. Therefore Nb$_2$S$_2$-2D should be chemisorbed to Gr. Given that the \ophase~islands are easy to move with the STM tip on Gr [see Figure~S2b in the SI] and that the C~1s core level is not affected during the phase transformations of physisorbed \nbs~[compare Figure~S4b in the SI], the \ophase~is not chemisorbed. Thus, we can clearly exclude that the Nb$_2$S$_2$-2D is observed in our experiments, regardless of the stacking sequence.

\begin{figure}[h!]
	\centering
	\includegraphics[width=0.45\textwidth]{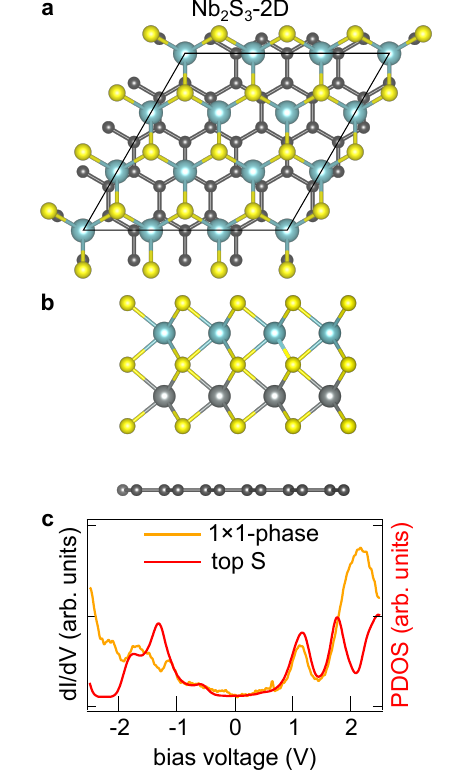}
	\caption{DFT calculated ball model representation of the \ophase~with stoichiometry Nb$_2$S$_3$-2D in (a) top and (b) side view. Nb atoms: cadet blue balls and gray; S atoms: yellow; dark gray: C atoms.
	(c) Differential conductance spectrum of the \ophase~(orange) compared to the DFT calculated partial density of states (PDOS) of the topmost S atoms (red).
	Spectrum parameters are $V_\mathrm{stab} = 2.5$\,V, $I_\mathrm{stab} = 0.7$\,nA, $V_\mathrm{mod} = 10$\,mV, 
	$f_\mathrm{mod} = 811$\,Hz, $T_\mathrm{s} = 1.7$\,K.}
	\label{ophase}
\end{figure}

As chemisorption has to be ruled out, we are forced to assume the presence of an additional passivating S layer, \textit{i.e.}, the \ophase~to be Nb$_2$S$_3$-2D. Using DFT, all possible eight stacking sequences for Nb$_2$S$_3$-2D were calculated (compare Table~S2 of the SI). Irrespective of the stacking, \nbss-2D is solely physisorbed to Gr, consistent with complete Nb passivation by S. The lowest energy structure is presented in Figure~\ref{ophase}a and b as top and side view ball model. It is again not the expected NiAs-type structure, but displays all Nb atoms in trigonal prismatic coordination with the S and Nb atoms of the respective layers sitting atop each other. The structure possesses no reconstruction, as required. It has a lattice parameter of $a=0.333$\,nm in decent agreement with the experimental value of $0.330$\,nm. The calculated height of $0.976$\,nm matches reasonably well with the STM measured apparent heights of $0.93$\,nm at $+1.00$\,V and $0.94$\,nm at $-1.00$\,V. Figure~\ref{ophase}c shows tm{good} agreement between the calculated top-layer S partial density of states with a large-range differential conductance STS spectrum. 

The interpretation of the S~2p core level components is now straightforward given the Nb$_2$S$_3$-2D stoichiometry: The S$_{\mathrm{btw}}$ component at 163.13\,eV strongly shifted by 2.48\,eV with respect to S$_{\mathrm{NbS2}}$ arises from S atoms located between two Nb planes, which provides a more electropositive environment compared to S atoms with Nb neighbors on only one side. This explains the significant binding energy shift, indicating a substantially altered chemical environment. The S$_{\mathrm{top-1}}$ component at 161.05\,eV shifted by 0.45\,eV with respect to S$_{\mathrm{NbS2}}$ corresponds to the S top plane of atoms, while the bottom S atoms, with two Nb and two S atomic planes above, do not contribute significantly to the intensity due to substantial damping. The bottom S intensity is presumably hidden in the S$_{\mathrm{top-1}}$ component.

The \sphase~is quite similar to the \ophase~in terms of symmetry, lattice parameter, apparent height (0.99\,nm), and the overall shape of the S 2p core-level spectra. However, it displays a $\sqrt{3} \times \sqrt{3}$ superstructure and contains only about $5/3$ Nb atoms per unit cell. From previous work \cite{Jellinek60} it is known that in the NiAs-type bulk structures of stoichiometry Nb$_{2-x}$S$_2$, Nb vacancies are present in every second Nb layer. 

\begin{figure*}[h!]
	\centering
	\includegraphics[width=0.9\textwidth]{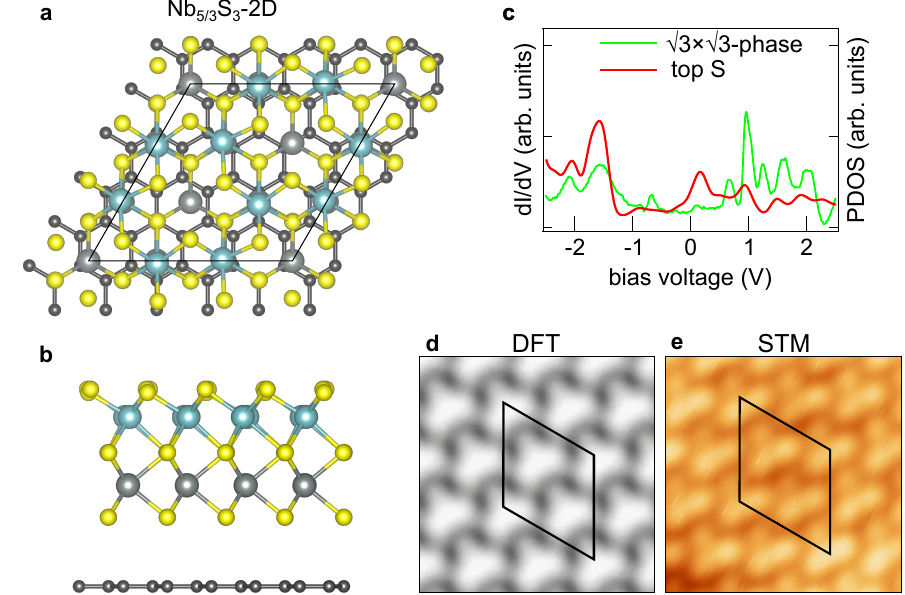}
 \caption{DFT calculated ball model representation of the \sphase~with stoichiometry Nb$_{5/3}$S$_3$-2D in (a) top and (b) side view. Nb atoms: cadet blue balls and gray; S atoms: yellow; dark gray: C atoms.
 (c) Differential conductance spectrum of the \sphase~(green) compared to the DFT calculated partial density of states (PDOS) of the topmost S atoms (red). (d) DFT calculated STM topograph compared to (e) measurement. Spectrum parameters: $V_\mathrm{stab} = 3$\,V, $I_\mathrm{stab} = 0.8$\,nA, $V_\mathrm{mod} = 10$\,mV, $f_\mathrm{mod} = 797$\,Hz, $T_\mathrm{s} = 1.7$\,K.}
\label{sphase}
\end{figure*}

For the \sphase~it is therefore most reasonable to assume that 1/3 of Nb is missing in one of the two Nb layers. This pattern would naturally give rise to the $\sqrt{3} \times \sqrt{3}$ superstructure, which is observed in LEED and STM. The stoichiometry of the \sphase~is consequently Nb$_{5/3}$S$_3$-2D.

The lowest energy structure of Nb$_{5/3}$S$_3$-2D is displayed in Figure~\ref{sphase}a and b as top and side view ball model (compare Table~S3 of the SI for other calculated structures). Again, the minimum energy is not a NiAs-type structure. While the Nb atoms in the complete atomic plane close to Gr are still in trigonal prismatic coordination, the top Nb plane with the regularly distributed Nb vacancies has the Nb atoms in octahedral coordination. The structure displays a $\sqrt{3} \times \sqrt{3}$ superstructure, as required. It has a lattice parameter of 0.333\,nm, in agreement with the experimental value of 0.333\,nm. The calculated height of 0.964\,nm matches reasonably well with the average of the STM measured apparent heights of 0.99\,nm at +1.00/\,V and 0.90\,nm at -1.00\,V. Figure~\ref{sphase}c shows decent agreement of the calculated top S partial density of states with a large-range differential conductance STS spectrum, reproducing the number of peaks on the unoccupied region, but not their location nor intensity. Figure~\ref{sphase}d and e show excellent agreement between the DFT simulated STM topograph and the measurement. 

The S 2p spectrum of the \sphase~is interpreted as follows: the S$_{\mathrm{top-\sqrt{3}}}$ component at 161.33 eV is assigned to top sulfur, while the S$_{\mathrm{btw}}$ component corresponds to the highly coordinated sulfur atoms between Nb atomic planes. The larger FWHM of both components for the \sphase, compared to the \ophase, indicates a less homogeneous state of sulfur, which due to the Nb vacancies, are bound to fewer Nb atoms on one side.

\section*{Summary}
\noindent
Under sulfur-poor conditions and heating single-layer \nbs~transforms to the more Nb-rich compounds Nb$_{5/3}$S$_3$-2D (1020\,K, \sphase) and Nb$_{2}$S$_3$-2D (1220\,K, \ophase). Nb$_{5/3}$S$_3$-2D displays a $\sqrt{3} \times \sqrt{3}$ superstructure caused by regularly arranged Nb vacancies in the top Nb layer. The same compounds may also be created by deposition of excess Nb under sulfur-poor conditions at 820\,K, a temperature at which \nbs~does not show changes with time in the absence of a Nb flux. The compounds consist of two Nb layers sandwiched between three S layers and are inert, covalently bound 2D materials. Consequently, these compounds emerge from \nbs~through covalent transformation. As uncovered by density functional theory calculations, the layer stacking sequence is unique for each compound and can not be derived from corresponding bulk materials. Their ease of preparation and unexpected structures provide a new handle to tailor properties and create new functions of 2D layers.

\section*{Methods}
\label{methods}
\noindent
The experiments were carried out in three ultrahigh vacuum systems (base pressure in low $10^{-10}$\,mbar range). All systems were equipped with sample preparation and growth facilities as well as LEED. STM measurements were conducted in two systems in Cologne while XPS was performed at the FlexPES beamline end station at MAX IV Laboratory, Lund. 

Ir(111) was cleaned by cycles of keV Ar$^+$ or Xe$^+$ sputtering and flash annealing to 1520\,K. Gr was grown by ethylene exposure of Ir(111) to saturation at room temperature, subsequent flash annealing to 1470\,K, and followed by exposure to $\approx 800$\,L of ethylene at 1370\,K. 
As confirmed by STM and LEED a closed single crystal Gr monolayer on Ir(111) results~\cite{vanGastel09}. 

Single-layer H-\nbs~was prepared by exposing Gr/Ir(111) to a flux of $\approx \mathrm{6\times10^{15}}$\,atoms/m$^2$s Nb from an e-beam evaporator in a background pressure of $\approx \mathrm{8\times 10^{-9}}$\,mbar elemental S. The S was supplied by a pyrite filled Knudsen cell about 10\,cm away from the sample. Growth was conducted for 510\,s at room temperature, followed by 360\,s annealing at 820\,K. During annealing, the Knudsen cell was turned off. However, since the S pressure decreases slowly, the S pressure remained non-zero, albeit well below $8 \times 10^{-9}$\,mbar. Since H-\nbs~growth takes place with excess S reevaporating, the amount of H-\nbs~formed is characterized through the amount of Nb deposited. 1 monolayer (ML) of Nb corresponds to the Nb amount in a full single layer of \nbs, \textit{i.e.}, to $1.12 \times 10^{19}$\,atoms/m$^2$. Phase transformations of  H-\nbs~resulted from deposition of elemental Nb onto single-layer \nbs~at different temperatures, or by annealing to temperatures above 820\,K, or both, as specified where the respective data is discussed. 

The samples were investigated \textit{in situ} by STM, either at 300\,K or at 1.7\,K after ultrahigh vacuum transfer into a bath cryostat. STS was conducted at 1.7\,K with Au-covered W tips calibrated using the surface state of Au(111) \cite{Kaiser86,Everson91}. Constant-current STM topographs were recorded with sample bias $V_\mathrm{s}$ and tunneling current $I_\mathrm{t}$ specified in each figure. $dI/dV$ spectra were recorded with stabilization bias $V_\mathrm{stab}$ and stabilization current $I_\mathrm{stab}$ using a lock-in amplifier with a modulation frequency $f_\mathrm{mod}$ and modulation voltage $V_\mathrm{mod}$, also specified in the captions.

The XPS experiments were conducted at the FlexPES beamline at
MAX IV Laboratory, Lund, Sweden \cite{Preobrajenski2023}. The growth of Nb$_x$S$_y$-2D compounds at the beamline was carried out with a Nb evaporator calibrated by STM in the home lab. High-resolution XPS of core-levels was performed in normal emission geometry with a spot size of $50~ \mu\mathrm{m} \times 50~\mu\mathrm{m}$ and at room temperature. The core levels were monitored with photon energies to maximize surface sensitivity: 150~eV for Ir~4f, 260~eV for S~2p, 380~eV for C~1s, 300~eV for Nb~3d. Overview spectra and high resolution O~1s spectra obtained at the first and last measurements of an annealing series confirmed that no other species were present. Curve fitting was performed with a pseudo-Voigt function. The asymmetry is included by an energy-dependent variation of the full-width-at-half maximum. The width, asymmetry and ratio of Gaussian to Lorentzian contributions were fixed for each component, meaning that they were not allowed to vary between spectra taken at different annealing temperatures. The center energy of each component was granted a $\pm100$~meV variation between different spectra while the intensities of the components were unconstrained. 

Our spin-polarized calculations were done by using DFT \cite{T1} and the projector augmented plane wave method \cite{T2} as implemented in the VASP code \cite{T3,T4}. We used a 500 eV energy cutoff for the plane wave expansion of the Kohn-Sham wave functions \cite{T5}. To account for the nonlocal correlation effects like van der Waals interactions \cite{T9}, all structural relaxations were done by using vdW-DF2 \cite{T6} functional containing a revised Becke (B86b) exchange \cite{T7,T8}, while the analysis of the electronic structure was performed by using the standard PBE exchange-correlation energy functional \cite{T10}.

\begin{acknowledgement}
We acknowledge funding from Deutsche Forschungsgemeinschaft (DFG) through CRC 1238 (project number 277146847, subprojects A01, B06 and C01). W.J. acknowledges financial support from the DFG SPP 2244 (projects No. 535290457). J.K. acknowledges financial support from the Swedish Research Council, grant number 2022-04363. J.F. acknowledges financial support from the DFG through project FI 2624/1-1 (project No. 462692705) within the SPP 2137. MAX IV Laboratory is acknowledged for time on Beamline FlexPES under Proposal 20210859. Research conducted at MAX IV, a Swedish national user facility, is supported by the Swedish Research council under contract 2018-07152, the Swedish Governmental Agency for Innovation Systems under contract 2018-04969, and Formas under contract 2019-02496. The authors acknowledge computing time granted by the JARA Vergabegremium and provided on the JARA Partition part of the supercomputer JURECA at Forschungszentrum Jülich.
\end{acknowledgement}

\begin{suppinfo}
Measurement of bilayer \nbs, Nb$_x$S$_y$ islands displaced by the STM tip, LEED of the \sphase, S~2p components as a function of temperature, Nb~3d, C~1s, and Ir~4f core-levels spectra as a function of temperature, DFT calculations of different Nb$_2$S$_2$-2D, Nb$_2$S$_3$-2D, and Nb$_{5/3}$S$_3$-2D structures on Gr.
\end{suppinfo}

\bibliography{bib_Nb2S3}

\providecommand{\latin}[1]{#1}
\makeatletter
\providecommand{\doi}
  {\begingroup\let\do\@makeother\dospecials
  \catcode`\{=1 \catcode`\}=2 \doi@aux}
\providecommand{\doi@aux}[1]{\endgroup\texttt{#1}}
\makeatother
\providecommand*\mcitethebibliography{\thebibliography}
\csname @ifundefined\endcsname{endmcitethebibliography}  {\let\endmcitethebibliography\endthebibliography}{}
\begin{mcitethebibliography}{42}
\providecommand*\natexlab[1]{#1}
\providecommand*\mciteSetBstSublistMode[1]{}
\providecommand*\mciteSetBstMaxWidthForm[2]{}
\providecommand*\mciteBstWouldAddEndPuncttrue
  {\def\EndOfBibitem{\unskip.}}
\providecommand*\mciteBstWouldAddEndPunctfalse
  {\let\EndOfBibitem\relax}
\providecommand*\mciteSetBstMidEndSepPunct[3]{}
\providecommand*\mciteSetBstSublistLabelBeginEnd[3]{}
\providecommand*\EndOfBibitem{}
\mciteSetBstSublistMode{f}
\mciteSetBstMaxWidthForm{subitem}{(\alph{mcitesubitemcount})}
\mciteSetBstSublistLabelBeginEnd
  {\mcitemaxwidthsubitemform\space}
  {\relax}
  {\relax}

\bibitem[Huang \latin{et~al.}(2015)Huang, Sutter, Shi, Zheng, Yang, Englund, Gao, and Sutter]{Huang15}
Huang,~Y.; Sutter,~E.; Shi,~N.~N.; Zheng,~J.; Yang,~T.; Englund,~D.; Gao,~H.-J.; Sutter,~P. Reliable exfoliation of large-area high-quality flakes of graphene and other two-dimensional materials. \emph{ACS Nano} \textbf{2015}, \emph{9}, 10612--10620\relax
\mciteBstWouldAddEndPuncttrue
\mciteSetBstMidEndSepPunct{\mcitedefaultmidpunct}
{\mcitedefaultendpunct}{\mcitedefaultseppunct}\relax
\EndOfBibitem
\bibitem[Huang \latin{et~al.}(2020)Huang, Pan, Yang, Bao, Meng, Luo, Cai, Liu, Zhao, Zhou, Wu, Zhu, Huang, Liu, Liu, Cheng, Wu, Tian, Gu, Shi, Guo, Cheng, Hu, Zhao, Yang, Sutter, Sutter, Wang, Ji, Zhou, and Gao]{Huang20}
Huang,~Y. \latin{et~al.}  {Universal mechanical exfoliation of large-area 2D crystals}. \emph{Nat. Commun.} \textbf{2020}, \emph{11}, 2453\relax
\mciteBstWouldAddEndPuncttrue
\mciteSetBstMidEndSepPunct{\mcitedefaultmidpunct}
{\mcitedefaultendpunct}{\mcitedefaultseppunct}\relax
\EndOfBibitem
\bibitem[Geim and Grigorieva(2013)Geim, and Grigorieva]{Geim13}
Geim,~A.~K.; Grigorieva,~I.~V. Van der {Waals} heterostructures. \emph{Nature} \textbf{2013}, \emph{499}, 419--425\relax
\mciteBstWouldAddEndPuncttrue
\mciteSetBstMidEndSepPunct{\mcitedefaultmidpunct}
{\mcitedefaultendpunct}{\mcitedefaultseppunct}\relax
\EndOfBibitem
\bibitem[Novoselov \latin{et~al.}(2016)Novoselov, Mishchenko, Carvalho, and Neto]{Novoselov16}
Novoselov,~K.~S.; Mishchenko,~A.; Carvalho,~A.; Neto,~A. H.~C. {2D materials and van der {Waals} heterostructures}. \emph{Science} \textbf{2016}, \emph{353}, aac9439\relax
\mciteBstWouldAddEndPuncttrue
\mciteSetBstMidEndSepPunct{\mcitedefaultmidpunct}
{\mcitedefaultendpunct}{\mcitedefaultseppunct}\relax
\EndOfBibitem
\bibitem[Cao \latin{et~al.}(2018)Cao, Fatemi, Demir, Fang, Tomarken, Luo, Sanchez-Yamagishi, Watanabe, Taniguchi, Kaxiras, Ashoori, and Jarillo-Herrero]{Cao18}
Cao,~Y.; Fatemi,~V.; Demir,~A.; Fang,~S.; Tomarken,~S.~L.; Luo,~J.~Y.; Sanchez-Yamagishi,~J.~D.; Watanabe,~K.; Taniguchi,~T.; Kaxiras,~E.; Ashoori,~R.~C.; Jarillo-Herrero,~P. Correlated insulator behaviour at half-filling in magic-angle graphene superlattices. \emph{Nature} \textbf{2018}, \emph{556}, 80--84\relax
\mciteBstWouldAddEndPuncttrue
\mciteSetBstMidEndSepPunct{\mcitedefaultmidpunct}
{\mcitedefaultendpunct}{\mcitedefaultseppunct}\relax
\EndOfBibitem
\bibitem[Wan \latin{et~al.}(2022)Wan, Wickramaratne, Dreher, Harsh, Mazin, and Ugeda]{Wan22}
Wan,~W.; Wickramaratne,~D.; Dreher,~P.; Harsh,~R.; Mazin,~I.~I.; Ugeda,~M.~M. Nontrivial Doping Evolution of Electronic Properties in {Ising}-Superconducting Alloys. \emph{Adv. Mater.} \textbf{2022}, \emph{34}, 2200492\relax
\mciteBstWouldAddEndPuncttrue
\mciteSetBstMidEndSepPunct{\mcitedefaultmidpunct}
{\mcitedefaultendpunct}{\mcitedefaultseppunct}\relax
\EndOfBibitem
\bibitem[Fortin-Desch{\^e}nes \latin{et~al.}(2024)Fortin-Desch{\^e}nes, Watanabe, Taniguchi, and Xia]{Fortin24}
Fortin-Desch{\^e}nes,~M.; Watanabe,~K.; Taniguchi,~T.; Xia,~F. Van der {Waals} epitaxy of tunable moir{\'e}s enabled by alloying. \emph{Nat. Mater.} \textbf{2024}, \emph{23}, 339--346\relax
\mciteBstWouldAddEndPuncttrue
\mciteSetBstMidEndSepPunct{\mcitedefaultmidpunct}
{\mcitedefaultendpunct}{\mcitedefaultseppunct}\relax
\EndOfBibitem
\bibitem[Zhao \latin{et~al.}(2020)Zhao, Song, Wang, Riis-Jensen, Fu, Deng, Wan, Kang, Ning, Dan, Venkatesan, Liu, Zhou, Thygesen, Luo, Pennycook, and Loh]{Zhao20}
Zhao,~X. \latin{et~al.}  {Engineering covalently bonded 2D layered materials by self-intercalation}. \emph{Nature} \textbf{2020}, \emph{581}, 171--177\relax
\mciteBstWouldAddEndPuncttrue
\mciteSetBstMidEndSepPunct{\mcitedefaultmidpunct}
{\mcitedefaultendpunct}{\mcitedefaultseppunct}\relax
\EndOfBibitem
\bibitem[Liu \latin{et~al.}(2021)Liu, Huang, Gou, Liang, Chua, Arramel, Duan, Zhang, Cai, Yu, Zhong, Zhang, and Wee]{Liu21}
Liu,~M.; Huang,~Y.~L.; Gou,~J.; Liang,~Q.; Chua,~R.; Arramel; Duan,~S.; Zhang,~L.; Cai,~L.; Yu,~X.; Zhong,~D.; Zhang,~W.; Wee,~A. T.~S. Diverse Structures and Magnetic Properties in Nonlayered Monolayer Chromium Selenide. \emph{J. Phys. Chem. Lett.} \textbf{2021}, \emph{12}, 7752--7760\relax
\mciteBstWouldAddEndPuncttrue
\mciteSetBstMidEndSepPunct{\mcitedefaultmidpunct}
{\mcitedefaultendpunct}{\mcitedefaultseppunct}\relax
\EndOfBibitem
\bibitem[Arnold \latin{et~al.}(2018)Arnold, Stan, Mahatha, Lund, Curcio, Dendzik, Bana, Travaglia, Bignardi, Lacovig, Lizzit, Li, Bianchi, Miwa, Bremholm, Lizzit, Hofmann, and Sanders]{Arnold18}
Arnold,~F. \latin{et~al.}  Novel single-layer vanadium sulphide phases. \emph{2D Mater.} \textbf{2018}, \emph{5}, 045009\relax
\mciteBstWouldAddEndPuncttrue
\mciteSetBstMidEndSepPunct{\mcitedefaultmidpunct}
{\mcitedefaultendpunct}{\mcitedefaultseppunct}\relax
\EndOfBibitem
\bibitem[van Efferen \latin{et~al.}(2024)van Efferen, Hall, Atodiresei, Boix, Safeer, Wekking, Vinogradov, Preobrajenski, Knudsen, Fischer, Jolie, and Michely]{Efferen24}
van Efferen,~C.; Hall,~J.; Atodiresei,~N.; Boix,~V.; Safeer,~A.; Wekking,~T.; Vinogradov,~N.~A.; Preobrajenski,~A.~B.; Knudsen,~J.; Fischer,~J.; Jolie,~W.; Michely,~T. {2D} vanadium sulfides: synthesis, atomic Structure engineering, and charge density Waves. \emph{ACS Nano} \textbf{2024}, \emph{18}, 14161--14175\relax
\mciteBstWouldAddEndPuncttrue
\mciteSetBstMidEndSepPunct{\mcitedefaultmidpunct}
{\mcitedefaultendpunct}{\mcitedefaultseppunct}\relax
\EndOfBibitem
\bibitem[Lasek \latin{et~al.}(2022)Lasek, Ghorbani-Asl, Pathirage, Krasheninnikov, and Batzill]{Lasek22}
Lasek,~K.; Ghorbani-Asl,~M.; Pathirage,~V.; Krasheninnikov,~A.~V.; Batzill,~M. {Controlling stoichiometry in ultrathin van der {Waals} Films: PtTe$_2$, Pt$_2$Te$_3$, Pt$_3$Te$_4$, and Pt$_2$Te$_2$}. \emph{ACS Nano} \textbf{2022}, \emph{16}, 9908--9919\relax
\mciteBstWouldAddEndPuncttrue
\mciteSetBstMidEndSepPunct{\mcitedefaultmidpunct}
{\mcitedefaultendpunct}{\mcitedefaultseppunct}\relax
\EndOfBibitem
\bibitem[Zhang \latin{et~al.}(2023)Zhang, Gong, Nie, Meng, Zhang, Gu, Liu, Lu, Fu, and Zhang]{Zhang23}
Zhang,~Z.-M.; Gong,~B.-C.; Nie,~J.-H.; Meng,~F.; Zhang,~Q.; Gu,~L.; Liu,~K.; Lu,~Z.-Y.; Fu,~Y.-S.; Zhang,~W. Self-intercalated {1T-FeSe$_2$} as an Effective Kagome Lattice. \emph{Nano Lett.} \textbf{2023}, \emph{23}, 954--961\relax
\mciteBstWouldAddEndPuncttrue
\mciteSetBstMidEndSepPunct{\mcitedefaultmidpunct}
{\mcitedefaultendpunct}{\mcitedefaultseppunct}\relax
\EndOfBibitem
\bibitem[Khatun \latin{et~al.}(2024)Khatun, Alanwoko, Pathirage, de~Oliveira, Tromer, Autreto, Galvao, and Batzill]{Khatun24}
Khatun,~S.; Alanwoko,~O.; Pathirage,~V.; de~Oliveira,~C.~C.; Tromer,~R.~M.; Autreto,~P. A.~S.; Galvao,~D.~S.; Batzill,~M. Solid State Reaction Epitaxy, A New Approach for Synthesizing Van der {Waals} heterolayers: The case of {Mn and Cr on Bi$_2$Se$_3$}. \emph{Adv. Funct. Mater.} \textbf{2024}, 2315112\relax
\mciteBstWouldAddEndPuncttrue
\mciteSetBstMidEndSepPunct{\mcitedefaultmidpunct}
{\mcitedefaultendpunct}{\mcitedefaultseppunct}\relax
\EndOfBibitem
\bibitem[{Van Maaren} and Schaeffer(1966){Van Maaren}, and Schaeffer]{vanmaaren66}
{Van Maaren},~M.~H.; Schaeffer,~G.~M. {Superconductivity in group Va dichalcogenides}. \emph{Phys. Lett.} \textbf{1966}, \emph{20}, 131\relax
\mciteBstWouldAddEndPuncttrue
\mciteSetBstMidEndSepPunct{\mcitedefaultmidpunct}
{\mcitedefaultendpunct}{\mcitedefaultseppunct}\relax
\EndOfBibitem
\bibitem[Fisher and Sienko(1980)Fisher, and Sienko]{Fisher80}
Fisher,~W.~G.; Sienko,~M.~J. {Stoichiometry, structure, and physical properties of niobium disulfide}. \emph{Inorganic Chemistry} \textbf{1980}, \emph{19}, 39--43\relax
\mciteBstWouldAddEndPuncttrue
\mciteSetBstMidEndSepPunct{\mcitedefaultmidpunct}
{\mcitedefaultendpunct}{\mcitedefaultseppunct}\relax
\EndOfBibitem
\bibitem[Witteveen \latin{et~al.}(2021)Witteveen, G{\'{o}}rnicka, Chang, M{\aa}nsson, Klimczuk, and von Rohr]{Witteven21}
Witteveen,~C.; G{\'{o}}rnicka,~K.; Chang,~J.; M{\aa}nsson,~M.; Klimczuk,~T.; von Rohr,~F.~O. {Polytypism and superconductivity in the NbS$_2$ system}. \emph{Dalton Trans.} \textbf{2021}, \emph{50}, 3216--3223\relax
\mciteBstWouldAddEndPuncttrue
\mciteSetBstMidEndSepPunct{\mcitedefaultmidpunct}
{\mcitedefaultendpunct}{\mcitedefaultseppunct}\relax
\EndOfBibitem
\bibitem[Lin \latin{et~al.}(2018)Lin, Huang, Zhao, Lian, Duan, Chen, and Ji]{Lin18}
Lin,~H.; Huang,~W.; Zhao,~K.; Lian,~C.; Duan,~W.; Chen,~X.; Ji,~S.~H. {Growth of atomically thick transition metal sulfide films on graphene/6H-SiC(0001) by molecular beam epitaxy}. \emph{Nano Res.} \textbf{2018}, \emph{11}, 4722--4727\relax
\mciteBstWouldAddEndPuncttrue
\mciteSetBstMidEndSepPunct{\mcitedefaultmidpunct}
{\mcitedefaultendpunct}{\mcitedefaultseppunct}\relax
\EndOfBibitem
\bibitem[Knispel \latin{et~al.}(2024)Knispel, Berges, Schobert, van Loon, Jolie, Wehling, Michely, and Fischer]{Knispel24}
Knispel,~T.; Berges,~J.; Schobert,~A.; van Loon,~E. G. C.~P.; Jolie,~W.; Wehling,~T.; Michely,~T.; Fischer,~J. {Unconventional Charge-Density-Wave Gap in Monolayer NbS$_2$}. \emph{Nano Lett.} \textbf{2024}, \emph{24}, 1045--1051\relax
\mciteBstWouldAddEndPuncttrue
\mciteSetBstMidEndSepPunct{\mcitedefaultmidpunct}
{\mcitedefaultendpunct}{\mcitedefaultseppunct}\relax
\EndOfBibitem
\bibitem[Yang \latin{et~al.}(2019)Yang, Mohmad, Wang, Fullon, Song, Zhao, Bozkurt, Augustin, Santos, Shin, Zhang, Voiry, Jeong, and Chhowalla]{Yang19}
Yang,~J.; Mohmad,~A.~R.; Wang,~Y.; Fullon,~R.; Song,~X.; Zhao,~F.; Bozkurt,~I.; Augustin,~M.; Santos,~E.~J.; Shin,~H.~S.; Zhang,~W.; Voiry,~D.; Jeong,~H.~Y.; Chhowalla,~M. {Ultrahigh-current-density niobium disulfide catalysts for hydrogen evolution}. \emph{Nat. Mater.} \textbf{2019}, \emph{18}, 1309--1314\relax
\mciteBstWouldAddEndPuncttrue
\mciteSetBstMidEndSepPunct{\mcitedefaultmidpunct}
{\mcitedefaultendpunct}{\mcitedefaultseppunct}\relax
\EndOfBibitem
\bibitem[Jellinek \latin{et~al.}(1960)Jellinek, Brauer, and M{\"{u}}ller]{Jellinek60}
Jellinek,~F.; Brauer,~G.; M{\"{u}}ller,~H. Molybdenum and Niobium Sulphides. \emph{Nature} \textbf{1960}, \emph{185}, 376--377\relax
\mciteBstWouldAddEndPuncttrue
\mciteSetBstMidEndSepPunct{\mcitedefaultmidpunct}
{\mcitedefaultendpunct}{\mcitedefaultseppunct}\relax
\EndOfBibitem
\bibitem[Kadijk and Jellinek(1969)Kadijk, and Jellinek]{Kadijk69}
Kadijk,~F.; Jellinek,~F. {The system niobium-sulfur}. \emph{J. Less-common metals} \textbf{1969}, \emph{19}, 421--430\relax
\mciteBstWouldAddEndPuncttrue
\mciteSetBstMidEndSepPunct{\mcitedefaultmidpunct}
{\mcitedefaultendpunct}{\mcitedefaultseppunct}\relax
\EndOfBibitem
\bibitem[Stan \latin{et~al.}(2019)Stan, Mahatha, Bianchi, Sanders, Curcio, Hofmann, and Miwa]{Stan19}
Stan,~R.-M.; Mahatha,~S.~K.; Bianchi,~M.; Sanders,~C.~E.; Curcio,~D.; Hofmann,~P.; Miwa,~J.~A. {Epitaxial single-layer ${\mathrm{NbS}}_{2}$ on Au(111): Synthesis, structure, and electronic properties}. \emph{Phys. Rev. Mater.} \textbf{2019}, \emph{3}, 44003\relax
\mciteBstWouldAddEndPuncttrue
\mciteSetBstMidEndSepPunct{\mcitedefaultmidpunct}
{\mcitedefaultendpunct}{\mcitedefaultseppunct}\relax
\EndOfBibitem
\bibitem[Schumacher \latin{et~al.}(2013)Schumacher, F\"orster, R\"osner, Wehling, and Michely]{Schumacher13}
Schumacher,~S.; F\"orster,~D.~F.; R\"osner,~M.; Wehling,~T.~O.; Michely,~T. Strain in Epitaxial Graphene Visualized by Intercalation. \emph{Phys. Rev. Lett.} \textbf{2013}, \emph{110}, 086111\relax
\mciteBstWouldAddEndPuncttrue
\mciteSetBstMidEndSepPunct{\mcitedefaultmidpunct}
{\mcitedefaultendpunct}{\mcitedefaultseppunct}\relax
\EndOfBibitem
\bibitem[Martínez-Galera \latin{et~al.}(2017)Martínez-Galera, Schröder, Herbig, Arman, Knudsen, and Michely]{Galera17}
Martínez-Galera,~A.~J.; Schröder,~U.~A.; Herbig,~C.; Arman,~M.~A.; Knudsen,~J.; Michely,~T. Preventing sintering of nanoclusters on graphene by radical adsorption. \emph{Nanoscale} \textbf{2017}, \emph{9}, 13618--13629\relax
\mciteBstWouldAddEndPuncttrue
\mciteSetBstMidEndSepPunct{\mcitedefaultmidpunct}
{\mcitedefaultendpunct}{\mcitedefaultseppunct}\relax
\EndOfBibitem
\bibitem[Hall \latin{et~al.}(2018)Hall, Pieli{\'{c}}, Murray, Jolie, Wekking, Busse, Kralj, and Michely]{Hall18}
Hall,~J.; Pieli{\'{c}},~B.; Murray,~C.; Jolie,~W.; Wekking,~T.; Busse,~C.; Kralj,~M.; Michely,~T. {Molecular beam epitaxy of quasi-freestanding transition metal disulphide monolayers on van der {Waals} substrates: a growth study}. \emph{2D Mater.} \textbf{2018}, \emph{5}, 25005\relax
\mciteBstWouldAddEndPuncttrue
\mciteSetBstMidEndSepPunct{\mcitedefaultmidpunct}
{\mcitedefaultendpunct}{\mcitedefaultseppunct}\relax
\EndOfBibitem
\bibitem[Pieli{\'{c}} \latin{et~al.}(2020)Pieli{\'{c}}, Hall, Despoja, Raki{\'{c}}, Petrovi{\'{c}}, Sohani, Busse, Michely, and Kralj]{Pielic20}
Pieli{\'{c}},~B.; Hall,~J.; Despoja,~V.; Raki{\'{c}},~I.~{\v{S}}.; Petrovi{\'{c}},~M.; Sohani,~A.; Busse,~C.; Michely,~T.; Kralj,~M. Sulfur Structures on Bare and Graphene-Covered {Ir(111)}. \emph{J. Phys. Chem. C} \textbf{2020}, \emph{124}, 6659--6668\relax
\mciteBstWouldAddEndPuncttrue
\mciteSetBstMidEndSepPunct{\mcitedefaultmidpunct}
{\mcitedefaultendpunct}{\mcitedefaultseppunct}\relax
\EndOfBibitem
\bibitem[van Gastel \latin{et~al.}(2009)van Gastel, N'Diaye, Wall, Coraux, Busse, Buckanie, zu~Heringdorf, von Hoegen, Michely, and Poelsema]{vanGastel09}
van Gastel,~R.; N'Diaye,~A.~T.; Wall,~D.; Coraux,~J.; Busse,~C.; Buckanie,~N.~M.; zu~Heringdorf,~F.-J.; von Hoegen,~M.; Michely,~T.; Poelsema,~B. {Selecting a single orientation for millimeter sized graphene sheets}. \emph{Appl. Phys. Lett.} \textbf{2009}, \emph{95}, 121901\relax
\mciteBstWouldAddEndPuncttrue
\mciteSetBstMidEndSepPunct{\mcitedefaultmidpunct}
{\mcitedefaultendpunct}{\mcitedefaultseppunct}\relax
\EndOfBibitem
\bibitem[Kaiser and Jaklevic(1986)Kaiser, and Jaklevic]{Kaiser86}
Kaiser,~W.~J.; Jaklevic,~R.~C. {Spectroscopy of electronic states of metals with a scanning tunneling microscope}. \emph{IBM J. Res. Dev.} \textbf{1986}, \emph{30}, 411--416\relax
\mciteBstWouldAddEndPuncttrue
\mciteSetBstMidEndSepPunct{\mcitedefaultmidpunct}
{\mcitedefaultendpunct}{\mcitedefaultseppunct}\relax
\EndOfBibitem
\bibitem[Everson(1991)]{Everson91}
Everson,~M.~P. {Effects of surface features upon the Au(111) surface state local density of states studied with scanning tunneling spectroscopy}. \emph{J. Vac. Sci. Technol.} \textbf{1991}, \emph{9}, 891\relax
\mciteBstWouldAddEndPuncttrue
\mciteSetBstMidEndSepPunct{\mcitedefaultmidpunct}
{\mcitedefaultendpunct}{\mcitedefaultseppunct}\relax
\EndOfBibitem
\bibitem[Preobrajenski \latin{et~al.}(2023)Preobrajenski, Generalov, {\"{O}}hrwall, Tchaplyguine, Tarawneh, Appelfeller, Frampton, and Walsh]{Preobrajenski2023}
Preobrajenski,~A.; Generalov,~A.; {\"{O}}hrwall,~G.; Tchaplyguine,~M.; Tarawneh,~H.; Appelfeller,~S.; Frampton,~E.; Walsh,~N. {FlexPES: a versatile soft X-ray beamline at MAXIV Laboratory}. \emph{J. Synchrotron Rad.} \textbf{2023}, \emph{30}, 831--840\relax
\mciteBstWouldAddEndPuncttrue
\mciteSetBstMidEndSepPunct{\mcitedefaultmidpunct}
{\mcitedefaultendpunct}{\mcitedefaultseppunct}\relax
\EndOfBibitem
\bibitem[Hohenberg and Kohn(1964)Hohenberg, and Kohn]{T1}
Hohenberg,~P.; Kohn,~W. Inhomogeneous electron gas. \emph{Phys. Rev.} \textbf{1964}, \emph{136}, B864\relax
\mciteBstWouldAddEndPuncttrue
\mciteSetBstMidEndSepPunct{\mcitedefaultmidpunct}
{\mcitedefaultendpunct}{\mcitedefaultseppunct}\relax
\EndOfBibitem
\bibitem[Bl\"ochl(1994)]{T2}
Bl\"ochl,~P.~E. Projector augmented-wave method. \emph{Phys. Rev. B} \textbf{1994}, \emph{50}, 17953\relax
\mciteBstWouldAddEndPuncttrue
\mciteSetBstMidEndSepPunct{\mcitedefaultmidpunct}
{\mcitedefaultendpunct}{\mcitedefaultseppunct}\relax
\EndOfBibitem
\bibitem[Kresse and Hafner(1993)Kresse, and Hafner]{T3}
Kresse,~G.; Hafner,~J. Ab initio molecular dynamics for liquid metals. \emph{Phys. Rev. B} \textbf{1993}, \emph{47}, 558\relax
\mciteBstWouldAddEndPuncttrue
\mciteSetBstMidEndSepPunct{\mcitedefaultmidpunct}
{\mcitedefaultendpunct}{\mcitedefaultseppunct}\relax
\EndOfBibitem
\bibitem[Kresse and Furthm\"uller(1996)Kresse, and Furthm\"uller]{T4}
Kresse,~G.; Furthm\"uller,~J. Ab initio molecular dynamics for liquid metals. \emph{Phys. Rev. B} \textbf{1996}, \emph{54}, 11169\relax
\mciteBstWouldAddEndPuncttrue
\mciteSetBstMidEndSepPunct{\mcitedefaultmidpunct}
{\mcitedefaultendpunct}{\mcitedefaultseppunct}\relax
\EndOfBibitem
\bibitem[Kohn and Sham(1965)Kohn, and Sham]{T5}
Kohn,~W.; Sham,~L.~J. Self-consistent equations including exchange and correlation effects. \emph{Phys. Rev.} \textbf{1965}, \emph{140}, A1133\relax
\mciteBstWouldAddEndPuncttrue
\mciteSetBstMidEndSepPunct{\mcitedefaultmidpunct}
{\mcitedefaultendpunct}{\mcitedefaultseppunct}\relax
\EndOfBibitem
\bibitem[Huttmann \latin{et~al.}(2015)Huttmann, Martinez~Galera, Caciuc, Atodiresei, Schumacher, Standop, Hamada, Wehling, Bl\"ugel, and Michely]{T9}
Huttmann,~F.; Martinez~Galera,~A.~J.; Caciuc,~V.; Atodiresei,~N.; Schumacher,~S.; Standop,~S.; Hamada,~I.; Wehling,~T.~O.; Bl\"ugel,~S.; Michely,~T. Tuning the van der {Waals} Interaction of Graphene with Molecules via Doping. \emph{Phys. Rev. Lett.} \textbf{2015}, \emph{115}, 236101\relax
\mciteBstWouldAddEndPuncttrue
\mciteSetBstMidEndSepPunct{\mcitedefaultmidpunct}
{\mcitedefaultendpunct}{\mcitedefaultseppunct}\relax
\EndOfBibitem
\bibitem[Lee \latin{et~al.}(2010)Lee, Murray, Kong, Lundqvist, and Langreth]{T6}
Lee,~K.; Murray,~E.~D.; Kong,~L.; Lundqvist,~B.~I.; Langreth,~D.~C. Higher accuracy van der {Waals} density functional. \emph{Phys. Rev. B} \textbf{2010}, \emph{82}, 081101\relax
\mciteBstWouldAddEndPuncttrue
\mciteSetBstMidEndSepPunct{\mcitedefaultmidpunct}
{\mcitedefaultendpunct}{\mcitedefaultseppunct}\relax
\EndOfBibitem
\bibitem[Becke(1986)]{T7}
Becke,~A. On the large-gradient behavior of the density functional exchange energy. \emph{J. Chem. Phys.} \textbf{1986}, \emph{85}, 7184\relax
\mciteBstWouldAddEndPuncttrue
\mciteSetBstMidEndSepPunct{\mcitedefaultmidpunct}
{\mcitedefaultendpunct}{\mcitedefaultseppunct}\relax
\EndOfBibitem
\bibitem[Hamada(2014)]{T8}
Hamada,~I. Higher-accuracy van der {Waals} density functional. \emph{Phys. Rev. B} \textbf{2014}, \emph{89}, 121103\relax
\mciteBstWouldAddEndPuncttrue
\mciteSetBstMidEndSepPunct{\mcitedefaultmidpunct}
{\mcitedefaultendpunct}{\mcitedefaultseppunct}\relax
\EndOfBibitem
\bibitem[Perdew \latin{et~al.}(1996)Perdew, Burke, and Ernzerhof]{T10}
Perdew,~J.~P.; Burke,~K.; Ernzerhof,~M. Generalized gradient approximation made simple. \emph{Phys. Rev. Lett.} \textbf{1996}, \emph{77}, 3865\relax
\mciteBstWouldAddEndPuncttrue
\mciteSetBstMidEndSepPunct{\mcitedefaultmidpunct}
{\mcitedefaultendpunct}{\mcitedefaultseppunct}\relax
\EndOfBibitem
\end{mcitethebibliography}

\end{document}


\DeclareGraphicsExtensions{.pdf}

\maketitle
\newpage
\tableofcontents
\newpage

\section*{\normalsize Supplementary Note 1: N\lowercase{b}S$_2$~bilayer islands after room temperature growth and annealing at 820~K}

\begin{figure*}[h!]
	\centering
	\includegraphics[width=0.9\textwidth]{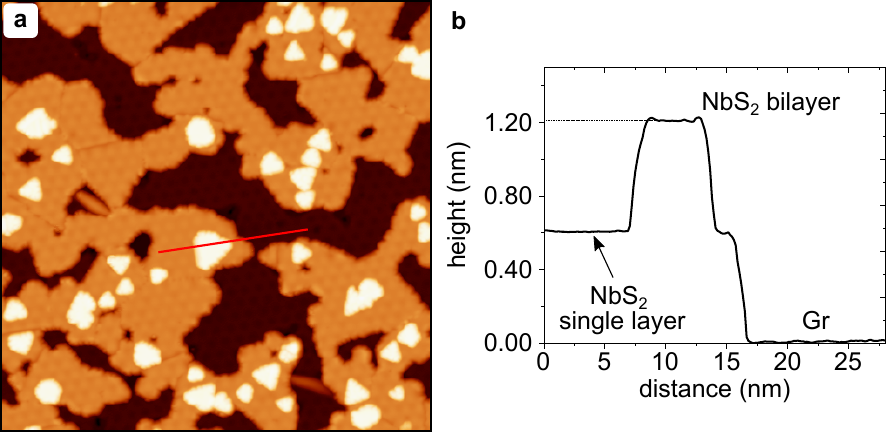}
	\caption{\nbs~single layer and bilayer islands after growth on Gr/Ir(111) and annealing to 820\,K. (a) STM topograph. (b) Height profile measured along the red line in (a). Image information: size $\mathrm{75\,nm\times75}$\,nm, $V_\mathrm{s}=1.0$\,V, $I_\mathrm{t}=0.1$\,nA, $T_\mathrm{s}=0.4$\,K} 
	\label{NbS2bilayer}
\end{figure*}

\newpage
\section*{\normalsize Supplementary Note 2: N\lowercase{b}$_x$S$_y$ islands displaced by the STM tip}
\begin{figure*}[h!]
	\centering
	\includegraphics[width=0.9\textwidth]{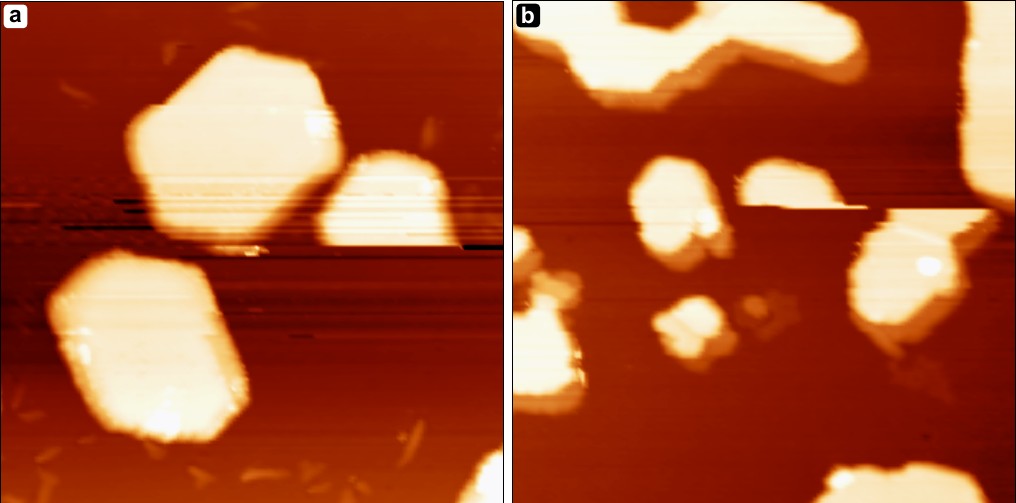}
	\caption{{(a)} Nb$_2$S$_3$-2D and (b) Nb$_{5/3}$S$_3$-2D~islands displaced by the STM tip under standard imaging conditions. Abrupt horizontal island cuts indicate displacement. The STM tip induced shift of both types of islands demonstrates their weak coupling to the substrate. Image information: (a,b) size $\mathrm{60\,nm\times}60$\,nm, $V_\mathrm{s}=1.0$\,V,  $I_\mathrm{t}=0.2$\,nA.}
	\label{scratched}
\end{figure*}

\newpage
\section*{\normalsize Supplementary Note 3: LEED of the \sphase}
\begin{figure}[h!]
	\centering
	\includegraphics[width=0.5\textwidth]{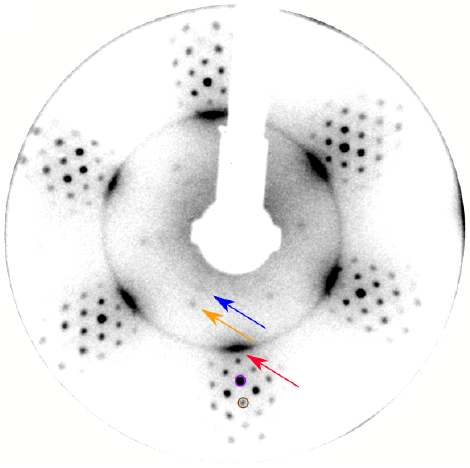}
	\caption{Inverted 140\,eV LEED pattern of the \sphase~obtained after annealing \nbs~to 1020~K. First order reflections of Ir and Gr are marked with magenta and brown circles, respectively. A fundamental spot of the \sphase~is marked by a red arrow. Two sets of $\sqrt3\times\sqrt3~\mathrm{R}30^\circ$ superstructure spots are observed. One is with respect to Ir and due to S intercalated between Ir(111) and Gr (green). The other set is with respect to the fundamental \sphase~ reflection (orange).}
	\label{LEED}
\end{figure}

\newpage
\section*{\normalsize Supplementary Note 4: S~2\lowercase{p} components and total S~2\lowercase{p} core level intensity as a function of temperature during \nbs~transformation by annealing.}
\begin{figure}[h]
	\centering
	\includegraphics[width=0.9\textwidth]{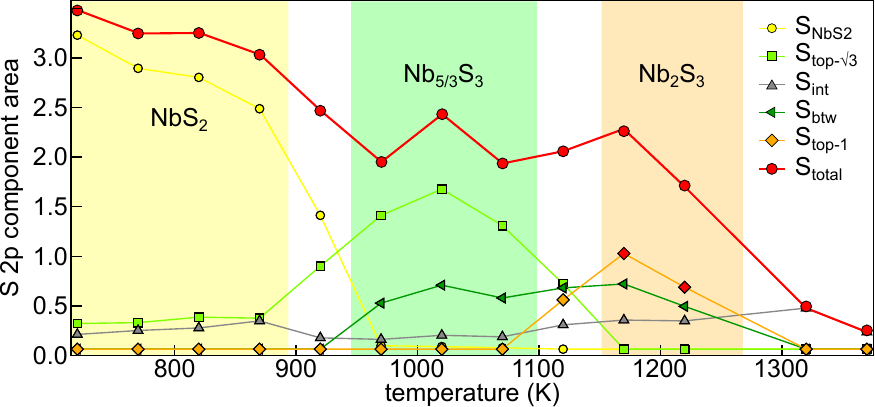}
	\caption{Integrated intensities of the fit components (as in Figure~4b,c,d in the main text) as a function of temperature. Total S~2p intensity in red. The background is color coded according to the dominating S~2p component: yellow (\nbs), green (\sphase), and orange (\ophase).
	}
	\label{xps_S2p_Nb3d}
\end{figure}

\newpage
\section*{\normalsize Supplementary Note 5: N\lowercase{b}~3\lowercase{d}, C~1\lowercase{s}, and I\lowercase{r}~4\lowercase{f} core-levels as a function of temperature during \nbs~transformation by annealing.}

\begin{figure}[h]
	\centering
	\includegraphics[width=0.9\textwidth]{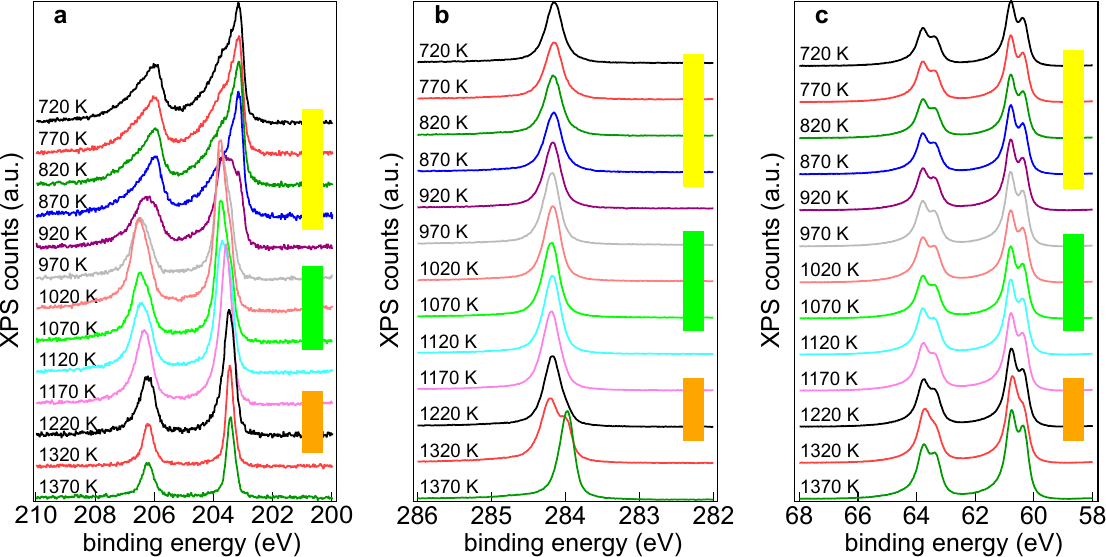}
	\caption{(a) High-resolution X-ray photoemission spectroscopy of the Nb~3d, C~1s, and Ir~4f core levels during \nbs~transformation by annealing from 720~K to 1370~K. Each data point was measured at room temperature after annealing to the indicated temperature without supply of addition S. Three temperature ranges are grouped according to the features identified in the S~2p core level from Figure~4a in the main text. The groups are color coded accordingly: yellow, green, and orange. Photon energies: $h\nu=300$~eV for Nb~3d, $h\nu=380$~eV for C~1s, and $h\nu=150$~eV for Ir~4f.
	}
	\label{xps_dissociation_Nb3d_C1s_Ir4f}
\end{figure}

\newpage
\section*{\normalsize Supplementary Note 6: DFT calculated  N\lowercase{b}$_2$S$_2$-2D structures on Gr}
\begin{table}[h!]
     \begin{center}
     \begin{tabular}{ | c | c | c | c | c | c | c|}
     \hline
     Nb coordination & 1H & 1H & 1T & 1T & 1H & 1H \\ \hline
     Nb-Nb position & shifted  & shifted & aligned & aligned & aligned & aligned  \\ \hline
     structure
     &
      \includegraphics[width=20mm]{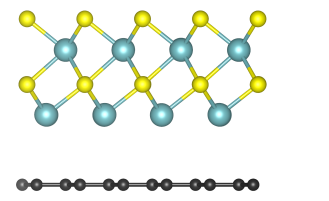}
      & 
	\includegraphics[width=20mm]{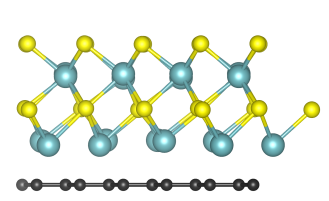}
      & 
	\includegraphics[width=20mm]{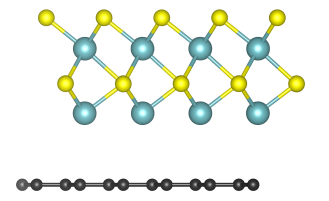}
      & 
	\includegraphics[width=20mm]{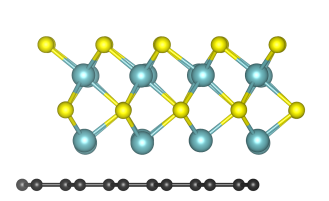}
      & 
	\includegraphics[width=20mm]{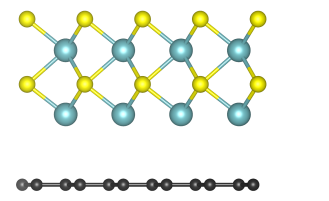}
      & 
	\includegraphics[width=20mm]{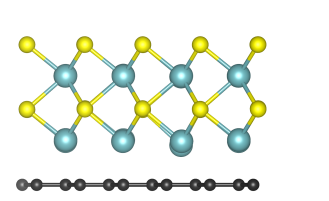}
      \\ \hline
     $\Delta E$ (eV/Nb$_2$S$_2$) & 1.552 & 0.719 & 0.952 & 0.198 & 0.766 & 0.0000  \\ \hline
	Nb-C bond& phys & chem & phys & chem & phys & chem  \\ \hline
     height (nm) & 0.827& 0.702 & 0.833 & 0.699 & 0.826 & 0.699  \\ \hline
     trimerization & $1\times1$ & yes & $1\times1$ & yes & $1\times1$ & $1\times1$  \\ \hline
      \end{tabular}
      \caption{DFT calculated Nb$_2$S$_2$-2D structures on Gr. Nb coordination is either trigonal prismatic (H) or octahedral (T)}
      \label{nb2s2_structures}
      \end{center}
      \end{table}

\newpage
\section*{\normalsize Supplementary Note 7: DFT calculated N\lowercase{b}$_2$S$_3$-2D structures  on Gr}

\begin{table}[h!]
     \begin{center}
     \begin{tabular}{| c | c | c | c | c |}
     \hline
     coordination & 1T/1T & 1H/1T & 1T/1H & 1H/1H  \\ \hline
     Nb-Nb position & aligned & aligned & aligned & aligned  \\ \hline
     structure
     & 
	\includegraphics[width=30mm]{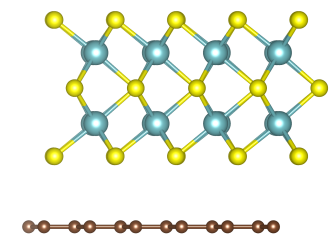}
      & 
	\includegraphics[width=30mm]{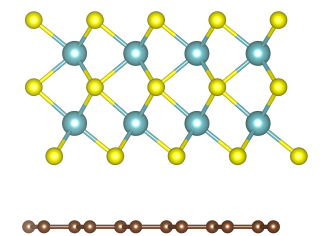}
      & 
	\includegraphics[width=30mm]{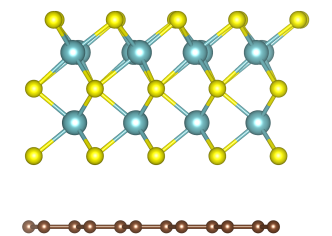}
      & 
	\includegraphics[width=30mm]{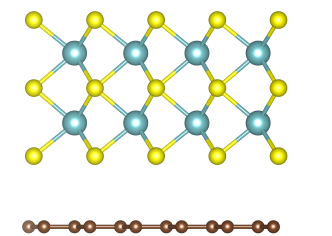}
      \\ \hline
     $\Delta E$ (eV/Nb$_2$S$_3$) & 0.4538 & 0.3049 & 0.2634 & 0.0000  \\ \hline
     height (nm) & 0.975 & 0.975 & 0.975 & 0.976  \\ \hline
     trimerization & yes & $1\times1$ & yes & $1\times1$  \\ \hline
      & & & &\\ \hline
     coordination & 1T/1T & 1T/1H & 1H/1T & 1H/1H \\ \hline
     Nb-Nb position & shifted & shifted & shifted & shifted  \\ \hline
     structure
     &
     \includegraphics[width=30mm]{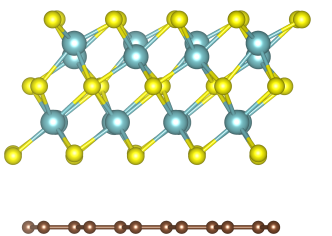}
      & 
	\includegraphics[width=30mm]{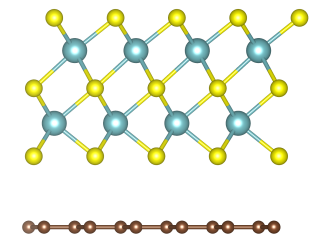}
     &
     \includegraphics[width=30mm]{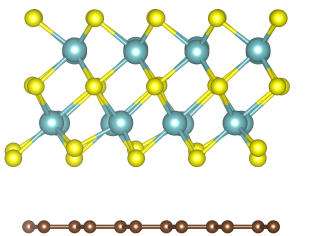}
      & 
	\includegraphics[width=30mm]{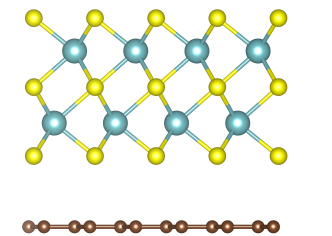}
      \\ \hline
     $\Delta E$ (eV/Nb$_2$S$_3$) & 1.3786 & 1.0962 & 1.0517 & 0.6858 \\ \hline
     height (nm) & 0.977 & 0.984 & 0.983& 0.984 \\ \hline
     trimerization & yes & $1\times1$ & yes & $1\times1$ \\ \hline
     \end{tabular}
      \caption{DFT calculated Nb$_2$S$_3$-2D structures on Gr for all possible stacking sequences. Nb coordination is either trigonal prismatic (H) or octahedral (T)}
      \label{nb2s3_structures}
      \end{center}
      \end{table}

\newpage
\section*{\normalsize Supplementary Note 8: 
DFT calculated N\lowercase{b}$_{5/3}$S$_3$-2D structures on Gr}

\begin{table}[h!]
    \begin{center}
     \begin{tabular}{| c | c | c | c | c | c | c | c |}
     \hline
      \multicolumn{2}{|c|}{coordination} & \multicolumn{2}{|c|}{1T/1H} & \multicolumn{2}{|c|}{1H/1H} & \multicolumn{2}{|c|}{1T/1T} \\ \hline
      \multicolumn{2}{|c|}{Nb-Nb position} & \multicolumn{2}{|c|}{aligned} & \multicolumn{2}{|c|}{aligned} & \multicolumn{2}{|c|}{aligned} \\ \hline
     \multicolumn{2}{|c|}{1/3 Nb missing} & ~top~ & bottom & ~top~ & bottom & ~top~ & bottom \\ \hline
     \multicolumn{2}{|c|}{structure}
    & 
	\includegraphics[width=20mm]{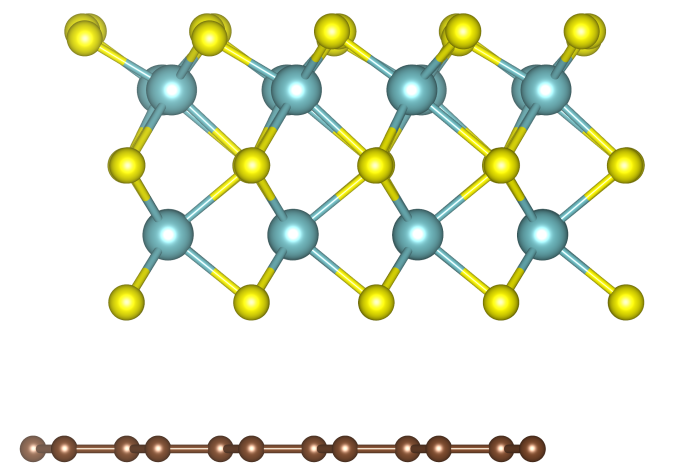}
    & 
	\includegraphics[width=20mm]{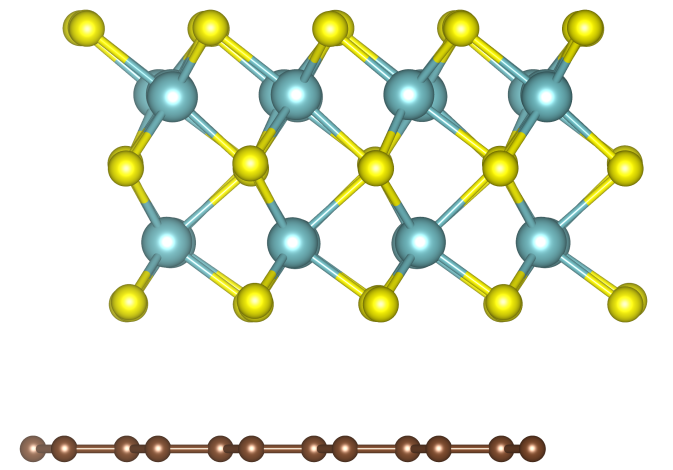}    
    & 
	\includegraphics[width=20mm]{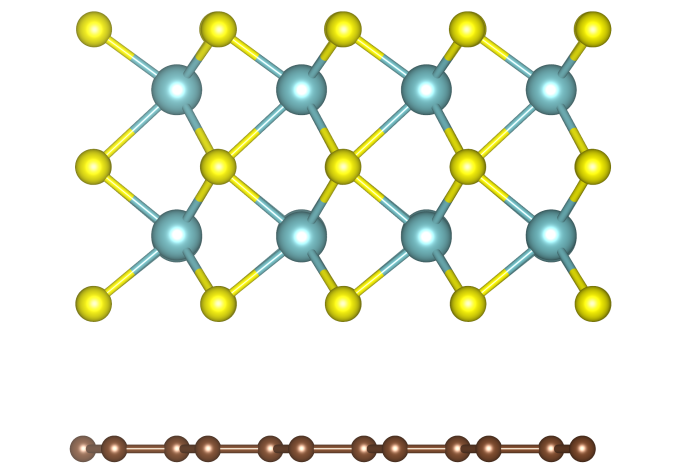}
    & 
	\includegraphics[width=20mm]{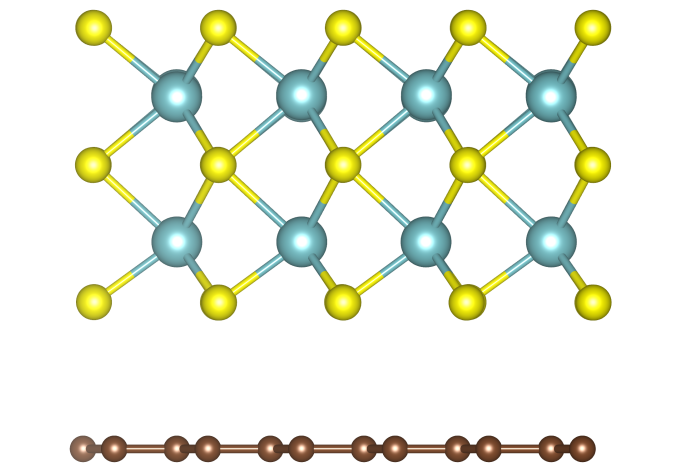}
    & 
	\includegraphics[width=20mm]{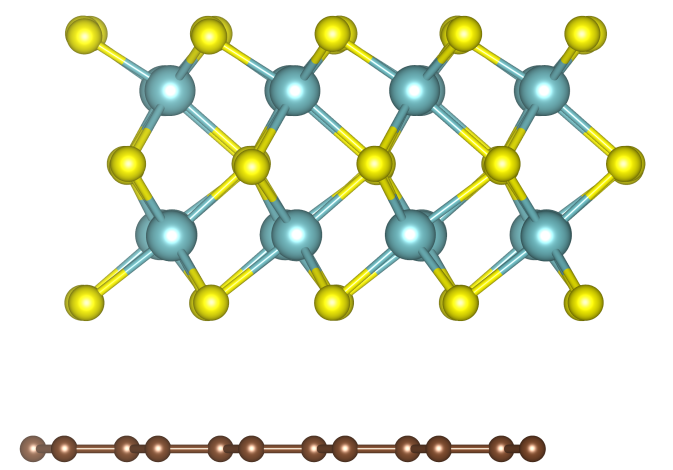}
    & 
	\includegraphics[width=20mm]{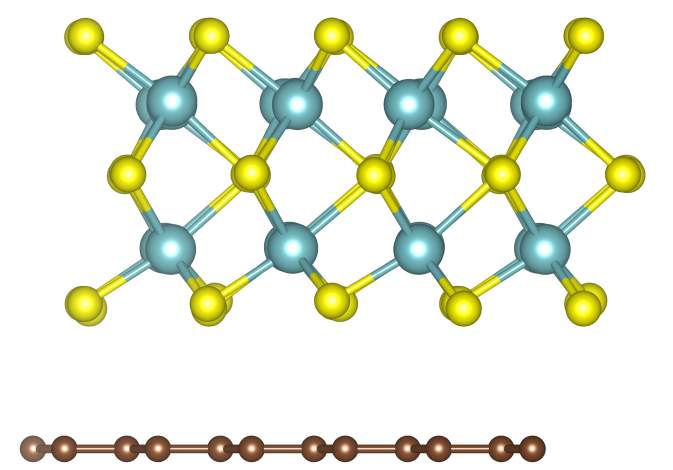}
    \\ \hline
     \multicolumn{2}{|c|}{height (nm)} & 0.964 & 0.970 & 0.968 & 0.972 & 0.958 & 0.954  \\ \hline
     \multicolumn{2}{|c|}{$\Delta E$ (eV/Nb$_{5/3}$S$_3$)} & ~~0.0000~~ & 0.6011 & ~~0.2378~~ & 0.2505 & ~~0.3736~~ & 0.3698  \\ \hline
      \end{tabular}
      \caption{Calculated Nb$_{5/3}$S$_3$-2D structures on Gr. Nb coordination is either trigonal prismatic (H) or octahedral (T)}
      \label{nb5/3s3_structures}
    \end{center}
\end{table}